\documentclass[12pt,english]{article}
\usepackage{times}
\usepackage[T1]{fontenc}
\usepackage{geometry}
\usepackage{graphicx}
\usepackage{setspace}
\usepackage{amssymb}

\makeatletter

\providecommand{\tabularnewline}{\\}

\usepackage{bm}

\usepackage{babel}
\makeatother
\begin{document}

\section*{\noindent A Divide-and-Conquer Tiling Method for the Design of Large
Aperiodic Phased Arrays}

\noindent \vfill

\noindent N. Anselmi,$^{(1)(2)}$ \emph{Senior Member}, \emph{IEEE},
P. Rocca,$^{(1)(2)(3)}$ \emph{Fellow}, \emph{IEEE}, G. Toso,$^{(4)}$
\emph{Fellow, IEEE}, and A. Massa,$^{(1)(2)(5)(6)(7)}$ \emph{Fellow,
IEEE}

\noindent \vfill

\noindent {\scriptsize ~}{\scriptsize \par}

\noindent {\scriptsize $^{(1)}$} \emph{\scriptsize ELEDIA Research
Center} {\scriptsize (}\emph{\scriptsize ELEDIA}{\scriptsize @}\emph{\scriptsize UniTN}
{\scriptsize - University of Trento)}{\scriptsize \par}

\noindent {\scriptsize DICAM - Department of Civil, Environmental,
and Mechanical Engineering}{\scriptsize \par}

\noindent {\scriptsize Via Mesiano 77, 38123 Trento - Italy}{\scriptsize \par}

\noindent \textit{\emph{\scriptsize E-mail:}} {\scriptsize \{}\emph{\scriptsize nicola.anselmi.1}{\scriptsize ,}
\emph{\scriptsize paolo.rocca}{\scriptsize ,} \emph{\scriptsize andrea.massa}{\scriptsize \}@}\emph{\scriptsize unitn.it}{\scriptsize \par}

\noindent {\scriptsize Website:} \emph{\scriptsize www.eledia.org/eledia-unitn}{\scriptsize \par}

\noindent {\scriptsize ~}{\scriptsize \par}

\noindent {\scriptsize $^{(2)}$} \emph{\scriptsize CNIT - \char`\"{}University
of Trento\char`\"{} ELEDIA Research Unit }{\scriptsize \par}

\noindent {\scriptsize Via Mesiano 77, 38123 Trento - Italy }{\scriptsize \par}

\noindent {\scriptsize E-mail: \{}\emph{\scriptsize nicola.anselmi}{\scriptsize ,}
\emph{\scriptsize paolo.rocca}{\scriptsize ,} \emph{\scriptsize andrea.massa}{\scriptsize \}}\emph{\scriptsize @unitn.it}{\scriptsize \par}

\noindent {\scriptsize Website:} \emph{\scriptsize www.eledia.org/eledia-unitn}{\scriptsize \par}

\noindent {\scriptsize ~}{\scriptsize \par}

\noindent {\scriptsize $^{(3)}$ ELEDIA Research Center (ELEDIA@XIDIAN
- Xidian University)}{\scriptsize \par}

\noindent {\scriptsize P.O. Box 191, No.2 South Tabai Road, 710071
Xi'an, Shaanxi Province - China}{\scriptsize \par}

\noindent {\scriptsize E-mail:} \emph{\scriptsize paolo.rocca@xidian.edu.cn}{\scriptsize \par}

\noindent {\scriptsize Website:} \emph{\scriptsize www.eledia.org/eledia-xidian}{\scriptsize \par}

\noindent {\scriptsize ~}{\scriptsize \par}

\noindent {\footnotesize $^{(4)}$} {\scriptsize Radio Frequency Payloads
and Technology Division, European Space Research and Technology Centre
(ESTEC), }{\scriptsize \par}

\noindent {\scriptsize European Space Agency (ESA), 2200AG Noordwijk,
The Netherland }{\scriptsize \par}

\noindent {\scriptsize E-mail:} \emph{\scriptsize giovanni.toso@esa.int}{\scriptsize \par}

\noindent {\scriptsize Website:} \emph{\scriptsize www.esa.int}{\scriptsize \par}

\noindent {\scriptsize ~}{\scriptsize \par}

\noindent {\scriptsize $^{(5)}$} \emph{\scriptsize ELEDIA Research
Center} {\scriptsize (}\emph{\scriptsize ELEDIA}{\scriptsize @}\emph{\scriptsize UESTC}
{\scriptsize - UESTC)}{\scriptsize \par}

\noindent {\scriptsize School of Electronic Science and Engineering,
Chengdu 611731 - China}{\scriptsize \par}

\noindent \textit{\emph{\scriptsize E-mail:}} \emph{\scriptsize andrea.massa@uestc.edu.cn}{\scriptsize \par}

\noindent {\scriptsize Website:} \emph{\scriptsize www.eledia.org/eledia}{\scriptsize -}\emph{\scriptsize uestc}{\scriptsize \par}

\noindent {\scriptsize ~}{\scriptsize \par}

\noindent {\scriptsize $^{(6)}$} \emph{\scriptsize ELEDIA Research
Center} {\scriptsize (}\emph{\scriptsize ELEDIA@TSINGHUA} {\scriptsize -
Tsinghua University)}{\scriptsize \par}

\noindent {\scriptsize 30 Shuangqing Rd, 100084 Haidian, Beijing -
China}{\scriptsize \par}

\noindent {\scriptsize E-mail:} \emph{\scriptsize andrea.massa@tsinghua.edu.cn}{\scriptsize \par}

\noindent {\scriptsize Website:} \emph{\scriptsize www.eledia.org/eledia-tsinghua}{\scriptsize \par}

\noindent {\scriptsize ~}{\scriptsize \par}

\noindent {\scriptsize $^{(7)}$} \emph{\scriptsize }{\scriptsize School
of Electrical Engineering}{\scriptsize \par}

\noindent {\scriptsize Tel Aviv University, Tel Aviv 69978 - Israel}{\scriptsize \par}

\noindent \textit{\emph{\scriptsize E-mail:}} \emph{\scriptsize andrea.massa@eng.tau.ac.il}{\scriptsize \par}

\noindent {\scriptsize Website:} \emph{\scriptsize https://engineering.tau.ac.il/}{\scriptsize \par}

\noindent \vfill

\newpage
\section*{A Divide-and-Conquer Tiling Method for the Design of Large Aperiodic
Phased Arrays}

~

\noindent ~

\noindent ~

\begin{flushleft}N. Anselmi, P. Rocca, G. Toso, and A. Massa\end{flushleft}

\vfill

\begin{abstract}
Due to the growing request from modern wireless applications of cost-affordable
and high-gain scanning antenna solutions, the design of large phased
arrays (\emph{PA}s) with radiating elements organized into modular
clusters with sub-array-only amplitude and phase control is a key
topic. In this paper, an innovative irregular tiling method is proposed
where, according to a divide-and-conquer strategy, the antenna aperture
is subdivided into sub-areas that are locally domino-tiled by jointly
fulfilling the full-coverage condition on the remaining untiled part
of the \emph{PA} support. Selected representative results, including
comparisons with competitive state-of-the-art synthesis methods, are
reported to prove the effectiveness and the computational efficiency
of the proposed tiling approach. Use-cases of current relevance for
low Earth orbit (\emph{LEO}) satellite communications are discussed,
as well, to provide the antenna designers useful practical guidelines
for handling large \emph{PA}s.

\noindent \vfill
\end{abstract}
\noindent \textbf{Key words}: Phased Array, Large Array Antenna, Sub-array,
Tiling, Domino, Optimization-based Synthesis, Satellite Communications.

\newpage
\section{Introduction}

\noindent Phased array (\emph{PA}) antennas are nowadays a widely
spread technology, but still complex and expensive for space-based
communication services based on low-orbit satellites. Indeed, antenna
solutions for fixed/mobile ground stations or in space satellites
\cite{Wang 2021} require wide beam steering angles and high gains,
while large apertures, filled by thousands of radiating elements,
imply too high implementation costs. An effective way to reduce the
\emph{PA} architecture complexity, the antenna weight, and the power
consumption is that of making the array layout highly modular so that
the antenna fabrication and its deployment become easier, scalable
\cite{Herd 2016}, and the number of amplifiers and phase shifters
is reduced, as well. Towards this end, unconventional \emph{PA}s (e.g.,
clustered layouts) have been deeply investigated \cite{Rocca 2016}
even though the reduction of the amplitude and phase controls generally
causes the appearance of grating lobes and the presence of high quantization
lobes, whose intensity and number grow with the average electrical
size of the clusters and the scan angle \cite{Mailloux 2009}-\cite{Haupt 2010}.
To cope with these problems, the aperiodic arrangement of the sub-array
phase centers turns out to be a reliable strategy as proven in the
seminal works \cite{Mailloux 2009}\cite{Vigano 2009}. Therefore,
several approaches to the synthesis of irregularly clustered arrays
have been proposed \cite{Xiong 2013}-\cite{Yang 2022} where polyomino
shaped tiles \cite{Xiong 2013}-\cite{Ma 2019}, tile panels characterized
by sparse \cite{Diao 2017} or clustered \cite{Li 2021} arrangements
of the array elements within the tiles, diamond shaped modules \cite{Rocca 2020a}
or different tiles sizes \cite{Rocca 2020b} have been used. 

\noindent In the recent literature, the most investigated polyomino
shape is the \emph{domino} one \cite{Anselmi 2017}-\cite{Ma 2019}\cite{Russo 1993}-\cite{Yang 2022}
where two elementary pixels are horizontally or vertically grouped
in a rectangular shape. A domino building block presents several positive
features such as the full coverage of rectangular as well as arbitrary
orthogonal-polygon like apertures \cite{Rocca 2022}, while affording
sub-array layouts suitable for an easy fabrication and assembly. Moreover,
domino tiling avoids half transmit-receive modules (\emph{TRM}s) of
a fully-populated (\emph{FP}) array still guaranteeing an adequate
clustering ratio for fulfilling challenging radiation requirements.
Furthermore, only two types of tiles, namely the vertical domino and
the horizontal one, are needed to fit user-desired polarization states,
unless the circular polarization case for which the dominoes can be
all equal \cite{Anselmi 2017}.

\noindent Dealing with domino clustering, customized tiling theorems
have been introduced to state the necessary conditions for the full
coverage of the antenna aperture in case of rectangular \cite{Anselmi 2017}
and arbitrary orthogonal-polygon \cite{Rocca 2022} arrays. Enumerative
\cite{Anselmi 2017} or optimization-driven \cite{Rocca 2022} techniques
have been developed for determining the best tiling when dealing with
small- or medium-size arrays, respectively. Moreover, domino-tiled
\emph{PA}s with time-dependent control points \cite{Ma 2019} have
been proposed, as well, or the maximization of the peak directivity
\cite{Yang 2021}\cite{Yang 2022} has been yielded by means of entropy-based
optimization techniques. However, all these synthesis methods suffer
either a slow convergence or too high computational costs when very
large arrays are at hand, which is the typical case of modern satellite
services where stringent regulatory pattern masks need to be satisfied
and high gains are required.

\noindent Despite the use of arbitrary polyominoes and without any
restriction on the number of copies of each tile shape, only small-
and medium-size \emph{PA}s have been optimally synthesized in \cite{Karademir 2015}
by means of an exact method based on a branch-and-bound strategy.
Otherwise, heuristic approaches have been implemented to handle wider
\emph{PA}s, but yielding sub-optimal tile arrangements.

\noindent This paper deals with the synthesis of large \emph{PA}s
with domino tiles by means of an innovative method based on a \char`\"{}divide
and conquer\char`\"{} strategy. Starting from a reference \emph{FP}
array layout, the antenna aperture is partitioned into multiple and
contiguous smaller sub-regions that are iteratively domino tiled,
in optimal way through analytic rules, by still fulfilling the full-coverage
(i.e., without holes) condition on the untiled remaining part of the
array \cite{Desreux 2006}. To enhance the irregularity of the arising
tilings, thus enabling enhanced radiation performance thanks to a
more exhaustive exploration of the solution space, soft boundaries
have been implemented among adjacent sub-regions to allow the inclusion
in a domino of two radiating elements not belonging to the same partition.

\noindent To the best of the authors' knowledge, the main novelties
of our research work over the existing literature \textbf{\emph{}}comprise
(\emph{i}) a customized method for the optimal tiling of large \emph{PA}s
using domino-shaped sub-arrays where, according to a divide-and-conquer
strategy, the antenna aperture of a reference \emph{FP} \emph{PA}
is subdivided into small and contiguous regions that are locally tiled,
while the radiation performance on the whole array layout are evaluated,
(\emph{ii}) a suitable use of analytic theorems for determining, sequentially
and exploiting soft-boundary concepts, the optimal tiling of each
sub-aperture by jointly fulfilling the full coverage condition for
the remaining untiled area of the \emph{PA} aperture, and (\emph{iii})
the derivation of engineering guidelines for the use of the proposed
method to handle large \emph{PA}s.

\noindent The rest of the paper is organized as follows. The mathematical
formulation of the synthesis of domino-tiled arrays is summarized
in Sect. \ref{sec:Mathematical-Formulation} by firstly defining the
tiling problem at hand (Sect. \ref{sub:Problem-Formulation}) and
then (Sect. \ref{sub:Divide-and-Conquer}) describing the proposed
{}``divide and conquer'' tiling method. Representative numerical
examples, including benchmark comparisons with competitive state-of-the-art
tiling methods and test cases of current practical relevance in \emph{LEO}
satcom applications, are reported in Sect. \ref{sec:Numerical-Assessment}.
Eventually, some conclusions and final remarks are drawn (Sect. \ref{sec:Conclusions}).

\section{Mathematical Formulation\label{sec:Mathematical-Formulation}}

\subsection{Problem Formulation\label{sub:Problem-Formulation}}

Let us consider a planar rectangular \emph{PA} with radiating elements
arranged on a rectangular lattice laying on the ($x,\, y$)-plane
, $d_{x}$ and $d_{y}$ being the inter-element spacing along the
$x$- and the $y$- axis, respectively (Fig. 1). Each element is assumed
to belong to a square unit cell, namely a \emph{pixel,}%
\footnote{\noindent For the sake of notation simplicity, each pixel is supposed
to include only one radiating element. However, each pixel is a logical
unit cell that can include one or more radiating elements without
loss of validity of the proposed tiling method.%
} such that the antenna aperture $\mathcal{A}$ turns out composed
of $M$ columns and $N$ rows, either $M$ or $N$ being even to fulfill
the full-coverage condition of $\mathcal{A}$ with domino tiles \cite{Anselmi 2017}.
The \emph{EM} field radiated in far-field by the array is given by\begin{eqnarray}
\mathbf{F}\left(u,v\right) & = & \sum_{m=1}^{M}\sum_{n=1}^{N}\mathbf{g}_{mn}\left(u,v\right)w_{mn}e^{j\frac{2\pi}{\lambda}\left(x_{m}u+y_{n}v\right)}\label{eq:_array.pattern}\end{eqnarray}
where $\mathbf{g}_{mn}\left(u,v\right)$ is the active/embedded element
pattern \cite{Mailloux 2018} of the $\left(m,n\right)$-th ($m=1,...,M$;
$n=1,...,N$) radiating element centered at $\left(x_{m},\, y_{n}\right)$
with complex excitation $w_{mn}$ ($w_{mn}=\alpha_{mn}e^{j\beta_{mn}}$,
$\alpha_{mn}$ and $\beta_{mn}$ being the corresponding amplitude
and phase coefficient, respectively). Moreover, $\lambda$ is the
wavelength at the antenna carrier frequency, $u=\sin\theta\cos\phi$
and $v=\sin\theta\sin\phi$ are the cosine angular directions, $\left(\theta,\phi\right)$
being the angular variables.

\noindent To simplify the array architecture, the $M\times N$ array
elements are clustered into $Q$ ($Q\triangleq\frac{M\times N}{2}$)
domino tiles so that each pixel belongs to a domino and the aperture
$\mathcal{A}$ is entirely covered. At the sub-array level, each tile
is connected to a single \emph{TRM} (Fig. 1) and the equivalent ($m,n$)-th
($m=1,...,M$; $n=1,...,N$) element-level excitation can be expressed
in terms of the $q$-th ($q=1,...,Q$) sub-array weight as follows

\noindent \begin{equation}
w_{mn}=\sum_{q=1}^{Q}\delta_{c_{mn}q}\alpha_{q}e^{j\beta_{q}}\label{eq:_equivalent.element-level.excitation}\end{equation}
($\widehat{\mathbf{w}}$ being the corresponding complex excitation
vector, $\widehat{\mathbf{w}}$ $\triangleq$ \{$w_{mn}$; $m=1,...,M$;
$n=1,...,N$\}) where $\bm{\alpha}=\left\{ \alpha_{q};\, q=1,...,Q\right\} $
and $\bm{\beta}=\left\{ \beta_{q};\, q=1,...,Q\right\} $ are the
real sub-array amplitude and phase $Q$-vectors, respectively, $\delta_{c_{mn}q}$
is the Kronecker delta function equal to $\delta_{c_{mn}q}=1$ when
the ($m,n$)-th ($m=1,...,M$; $n=1,...,N$) element belongs to the
$q$-th ($q=1,...,Q$) domino ($c_{mn}=q$), while $\delta_{c_{mn}q}=0$
otherwise ($c_{mn}\neq q$), $\mathbf{c}$ $=$ \{$c_{mn}$; $m=1,...,M$;
$n=1,...,N$\} being the array partitioning vector encoding the aggregation
of the $M\times N$ array elements into the $Q$ dominoes. 

\noindent The synthesis of domino-tiled \emph{PA}s can be then stated
as follows:

\begin{quote}
\noindent \emph{Domino-Tiled PA Synthesis} - Given an array aperture
$\mathcal{A}$ of $M\times N$ elements, determine the optimal clustering,
$\mathbf{c}^{opt}$, of the array elements into $Q$ domino sub-arrays
and the sets of sub-array amplitudes, $\bm{\alpha}^{opt}$, and phases,
$\bm{\beta}^{opt}$, that minimize the \emph{pattern matching} function\begin{equation}
\begin{array}{r}
\Phi\left(\mathbf{c}\right)\triangleq\frac{\int_{\Omega}\left[\mathcal{P}\left(u,v\right)-\Psi\left(u,v\right)\right]\times\mathcal{H}\left\{ \mathcal{P}\left(u,v\right)-\Psi\left(u,v\right)\right\} dudv}{\int_{\Omega}\Psi\left(u,v\right)dudv}\end{array}\label{eq:_cost.function}\end{equation}
where $\mathcal{P}\left(u,v\right)$ ($\mathcal{P}\left(u,v\right)\triangleq\left|\mathbf{F}\left(u,v\right)\right|^{2}$)
is the power pattern radiated by the domino-tiled \emph{PA}, $\Psi\left(u,v\right)$
is a positive upper-bound mask defined by the user over the angular
domain $\Omega$ ($\Omega$ $\triangleq$ \{$\left(u,v\right)$ |
$u^{2}+v^{2}\le1$\}), and $\mathcal{H}$ is the Heaviside step function.
\end{quote}
\noindent Such a synthesis problem is handled in Sect. \ref{sub:Divide-and-Conquer}
by means of an innovative method suitable for large \emph{PA}s, as
well.

\subsection{Divide and Conquer Tiling Method\label{sub:Divide-and-Conquer}}

\noindent The synthesis of domino-tiled \emph{PA}s has been already
addressed in \cite{Anselmi 2017} by means of an exhaustive tiling
strategy, named enumerative tiling method (\emph{ETM}), able to find
the optimal solution since iteratively exploring the whole solution
space of the $T$ admissible clustered configurations, $\mathbf{C}$
$=$ \{$\mathbf{c}^{\left(t\right)}$; $t=1,...,T$\}. Unfortunately,
the \emph{ETM} is realistically applicable only to small arrays because
of its computational cost. Therefore, the optimization-based tiling
method (\emph{OTM}) has been proposed in \cite{Anselmi 2017} to deal
with the synthesis of wider arrays thanks to the effective sampling
of the solution space yielded by an \emph{ad-hoc} method based on
a schemata-driven Genetic Algorithm (\emph{GA}). However, the synthesis
of bigger domino-tiled \emph{PA}s, including thousands of radiating
elements, turns out to be unaffordable for the \emph{OTM}, as well.
To overcome the dimensionality limitation of current tiling techniques,
we propose the divide-and-conquer tiling method (\emph{DCTM}), which
is aimed at providing suitable trade-offs between the optimality of
the synthesized tiling and the required computational burden. Towards
this end, the antenna aperture $\mathcal{A}$ of a reference \emph{FP-PA}
is first partitioned into $I$ smaller sub-apertures, \{$\mathcal{S}^{(i)}$;
$i=1,...,I$\} such that $\mathcal{A}=\bigcup_{i=1}^{I}\mathcal{S}^{(i)}$
and $\mathcal{S}^{(i)}\cap\mathcal{S}^{(j)}=0$ ($i,j=1,...,I$; $i\neq j$).
Then, each $i$-th ($i=1,...,I$) array sub-region, $\mathcal{S}^{(i)}$,
is progressively tiled using dominoes and assuming soft-boundaries
with the adjacent partitions to build dominoes with two close radiating
elements, but not exactly laying in the same $\mathcal{S}^{(i)}$.

\noindent More in detail, the \emph{DCTM} is implemented through the
sequence of the following steps:

\subsubsection*{Step 1 - Array Aperture Setup and Reference Pattern Selection}

\begin{itemize}
\item \textbf{Step 1.1} - \emph{Array Aperture Setup}. Let the pixels of
the aperture $\mathcal{A}$ be mapped in the checkerboard pattern
shown in Fig. 2(\emph{a}). By grouping the set of external and internal
vertices into the vectors $\bar{\mathbf{v}}=\left\{ \bar{v}_{b};\, b=1,...,B\right\} $
{[}$B\triangleq2\times\left(M+N\right)${]} ($\bar{v}_{b}\in\partial\mathcal{A}$,
$\partial\mathcal{A}$ being the periphery of $\mathcal{A}$) and
$\mathbf{v}=\left\{ v_{l};\, l=1,...,L\right\} $ {[}$L\triangleq\left(M-1\right)\times\left(N-1\right)${]},
respectively, let $\left(v_{l},\, v_{r}\right)$ be a generic edge
connecting the vertex $v_{l}$ to the vertex $v_{r}$ and let us assume
the edges of the white(grey) pixels oriented clockwise(counter-clockwise)
{[}Fig. 2(a){]};
\item \textbf{Step 1.2} - \emph{Height Function Computation}. For each $b$-th
($b=1,...,B$) external vertex, $\bar{v}_{b}$, compute the corresponding
value of the height function (\emph{HF}), $\bar{h}_{b}$ ($\bar{h}_{b}\triangleq h\left(\bar{v}_{b}\right)$
$\to$ $\bar{\mathbf{h}}=\left\{ \bar{h}_{b};\, b=1,...,B\right\} $)
\cite{Thurston 1990} by nullifying the \emph{HF} value of the first
($b=1$) vertex on the top left corner of the aperture ($\left.\bar{h}_{b}\right|_{b=1}=0$)
and then determining the \emph{HF} values of the remaining $B-1$
vertices ($b=2,...,B$) as follows\begin{equation}
\bar{h}_{b}=\left\{ \begin{array}{ll}
\bar{h}_{b-1}+1 & \;\;\; if\;\exists\;\left(\bar{v}_{b-1},\,\bar{v}_{b}\right)\\
\bar{h}_{b-1}-1 & \;\;\; if\;\exists\;\left(\bar{v}_{b},\,\bar{v}_{b-1}\right)\end{array}\right.\label{eq:_HF.external.vertices}\end{equation}
{[}Fig. 2(\emph{b}){]};
\item \textbf{Step 1.3} - \emph{Minimal Tiling Definition}. Define the minimal
tiling solution, $\left.\mathbf{c}^{\left(t\right)}\right|_{t=1}$,
and the associated \emph{HF} values of the internal vertices, $\mathbf{h}^{(1)}=\mathbf{h}\left(\mathbf{c}^{\left(1\right)}\right)$
{[}$\mathbf{h}=\left\{ h_{l};\, l=1,...,L\right\} $, $h_{l}\triangleq h\left(v_{l}\right)${]}
{[}Fig. 2(\emph{b}){]} according to the \emph{Algorithm B1} in \cite{Anselmi 2017};
\item \textbf{Step 1.4} - \emph{Aperture Partitioning}. Split the aperture
$\mathcal{A}$ into $I$ partitions, \{$\mathcal{S}^{(i)}$; $i=1,...,I$\},
of equal size $\widehat{M}\times\widehat{N}$ ($\widehat{M}\leq M$
and $\widehat{N}\leq N$) such that $\left(M\, mod\,\widehat{M}\right)=0$
and $\left(N\, mod\,\widehat{N}\right)=0$, $mod$ being the modulo
operation (e.g., Fig. 3 - $M=N=12$, $\widehat{M}=\widehat{N}=4$,
$I=9$). The coordinates of the ($r$, $s$)-th ($r=1,...,\widehat{M}$;
$s=1,...,\widehat{N}$) pixel of the $i$-th ($i=1,...,I$) sub-aperture,
$\mathcal{S}^{(i)}$, are ($x_{\widehat{m}_{r}}$, $y_{\widehat{n}_{s}}$)
where $\widehat{m}_{r}=r+\left(i-1-\frac{\left\lfloor \left(i-1\right)\times\frac{\widehat{M}}{M}\right\rfloor }{\frac{\widehat{M}}{M}}\right)\times\widehat{M}$
and $\widehat{n}_{s}=s+\left\lfloor \left(i-1\right)\times\frac{\widehat{M}}{M}\right\rfloor \times\widehat{N}$,
$\left\lfloor \cdot\right\rfloor $ being the floor function;
\item \textbf{Step 1.5} - \emph{Reference Pattern} \emph{Selection}. Define
the set of $M\times N$ amplitude, $\bm{\alpha}^{ref}=\left\{ \alpha_{mn}^{ref};\, m=1,...,M;\, n=1,...,N\right\} $,
and phase, $\bm{\beta}^{ref}=\left\{ \beta_{mn}^{ref};\, m=1,...,M;\, n=1,...,N\right\} $,
coefficients ($\to$ $\mathbf{w}^{ref}$ $=$ \{$w_{mn}^{ref}$; $m=1,...,M$;
$n=1,...,N$\}, $w_{mn}^{ref}\triangleq\alpha_{mn}^{ref}e^{j\beta_{mn}^{ref}}$)
of the \emph{FP} array affording a reference pattern, $\mathcal{P}^{ref}\left(u,v\right)$,
compliant with the user-defined requirements coded by $\Psi\left(u,v\right)$;
\end{itemize}

\subsubsection*{Step 2 - Divide-and-Conquer Domino Tiling Optimization}

\begin{itemize}
\item \textbf{Step 2.1} - \emph{Initialization}. At the first ($i=1$) iteration
of the \emph{DCTM}, set $\left.\mathcal{R}^{(i)}\right\rfloor _{i=1}=\mathcal{A}$
{[}$\mathcal{R}^{(i)}$ ($i=1,...,I-1$) being the untiled area of
$\mathcal{A}$, $\mathcal{R}^{(i)}\triangleq\mathcal{A}-\sum_{j=1}^{i-1}\mathcal{S}^{(j)}${]}
and $\widehat{\mathbf{w}}=\emptyset$ since the whole aperture has
to be tiled and no domino clusters have been used yet, respectively;
\item \textbf{Step 2.2} - $\mathcal{S}^{(i)}$ \emph{Sub-Aperture Tiling}.
To generate tiled configurations not only confined to $\mathcal{S}^{(i)}$,
but with dominoes that can potentially exceed the boundary between
$\mathcal{S}^{(i)}$ and the remaining part of $\mathcal{A}$ to be
processed, $\mathcal{D}^{(i)}$ ($\mathcal{D}^{(i)}\triangleq\mathcal{A}-\sum_{j=1}^{i}\mathcal{S}^{(j)}$
$\to$ $\mathcal{D}^{(i)}=\mathcal{R}^{(i)}-\mathcal{S}^{(i)}$) {[}light
yellow region in Fig. 4 ($i=1)$ and Fig. 5 ($i=2$){]}, consider
not only the internal vertices of $\mathcal{S}^{(i)}$, but also those
laying on the boundary between $\mathcal{S}^{(i)}$ and $\mathcal{D}^{(i)}$,
$\widehat{\mathbf{v}}=\left\{ v_{\widehat{l}};\,\widehat{l}=1,...,L^{(i)}\right\} $
{[}blue color dots in Fig. 4 ($i=1)$ and Fig. 5 ($i=2$){]}. Then,
if the condition $\eta\leq\eta_{th}$ ($\eta\triangleq\sqrt{\frac{\widehat{M}\times\widehat{N}}{M\times N}}$
and $\eta_{th}$ being a user-defined switching threshold) holds true,
use the \emph{enumerative DCTM} (\emph{E-DCTM}) strategy:

\begin{itemize}
\item Assign to the vertices $\widehat{\mathbf{v}}$, the \emph{HF} values,
$\widehat{\mathbf{h}}^{\left(1\right)}=\left\{ h_{\widehat{l}}^{\left(1\right)};\,\widehat{l}=1,...,L^{(i)}\right\} $,
of the minimal tiling, $\mathbf{h}^{\left(1\right)}$ (i.e., $h_{\widehat{l}}^{\left(1\right)}=h^{\left(1\right)}\left(v_{l\to\widehat{l}}\right)$
($\widehat{l}=1,...,L^{(i)}$; $l=1,...,L$), where the notation $l\to\widehat{l}$
indicates the $l$-th pixel in $\mathcal{A}$ that corresponds to
the $\widehat{l}$-th one in $\mathcal{S}^{(i)}$); 
\item Exhaustively generate all (i.e., including the layouts with dominoes
overlapping the soft boundaries of $\mathcal{S}^{(i)}$, as well)
tilings of $\mathcal{S}^{(i)}$, $\widehat{\mathbf{C}}$ = \{$\mathbf{c}^{\left(\widehat{t}\right)}$;
$\widehat{t}=1,...,T^{(i)}$\} according to the \emph{ETM} method
\cite{Anselmi 2017};
\item Check the admissibility of each $\widehat{t}$-th ($\widehat{t}=1,...,T^{(i)}$)
trial domino arrangement $\mathbf{c}^{\left(\widehat{t}\right)}$
by verifying the {}``\emph{Admissibility Condition}'', that is,
whether $\mathcal{R}^{(i+1)}$ can be fully covered with dominoes
according to the procedure developed for orthogonal polygon-shaped
apertures (Sect. 2.A \emph{}- \cite{Rocca 2022}). If $\mathcal{R}^{(i+1)}$
is not tileable and $\widehat{t}\ne T^{(i)}$, skip to the next clustering
($\widehat{t}\to\widehat{t}+1$), otherwise (i.e., $\mathcal{R}^{(i+1)}$
turns out to be tileable) set $\hat{\mathbf{c}}\leftarrow\mathbf{c}^{\left(\widehat{t}\right)}$
and determine the sub-array excitation vector $\widehat{\mathbf{w}}$
by computing the $q$-th ($q=1,...,Q^{(i)}$) entry of $\hat{\bm{\alpha}}$($\hat{\bm{\beta}}$)
as the average of the reference \emph{FP} amplitude(phase) coefficients
belonging to the $Q^{(i)}$ dominoes placed within $\mathcal{S}^{(i)}$\begin{equation}
\left(\begin{array}{c}
\hat{\alpha}_{q}\\
\hat{\beta}_{q}\end{array}\right)=\frac{1}{2}\sum_{r=1}^{\widehat{M}+1}\sum_{s=1}^{\widehat{N}+1}\left(\begin{array}{c}
\alpha_{\widehat{m}_{r}\widehat{n}_{s}}^{ref}\\
\beta_{\widehat{m}_{r}\widehat{n}_{s}}^{ref}\end{array}\right)\delta_{c_{\widehat{m}_{r}\widehat{n}_{s}}q}.\label{eq:_sub-array.amplitudes-phases_EM}\end{equation}
Update the equivalent element-level excitations as follows\begin{equation}
w_{mn}=\left\{ \begin{array}{ll}
\sum_{q=1}^{Q^{(i)}}\hat{\alpha}_{q}e^{j\hat{\beta}_{q}}\delta_{c_{mn}q} & \; if\;\left(x_{m},y_{n}\right)\in\mathcal{S}^{(i)}\\
\alpha_{mn}^{ref}e^{j\beta_{mn}^{ref}} & \; if\;\left(x_{m},y_{n}\right)\in\mathcal{R}^{(i+1)}\end{array}\right.\label{eq:_hybrid_array}\end{equation}
($m=1,...,M;\, n=1,...,N$). Given $\hat{\mathbf{c}}$, $\hat{\bm{\alpha}}$,
and $\hat{\bm{\beta}}$, compute the value of $\widehat{\Phi}$ {[}$\widehat{\Phi}\triangleq\Phi\left(\hat{\mathbf{c}}\right)=\Phi\left(\mathbf{c}^{\left(\widehat{t}\right)}\right)${]}
with (\ref{eq:_cost.function});
\end{itemize}
Otherwise, run the \emph{Optimization-Based DCTM} (\emph{O-DCTM})
technique:

\begin{itemize}
\item Set the initial ($k=0$, $k$ being the iteration index) $P$-size
population, $\widehat{\mathbf{C}}^{(k)}$ = \{$\mathbf{c}_{p}^{\left(k\right)}$;
$p=1,...,P$\}, according to the schemata-driven strategy in \cite{Anselmi 2017};
\item Iteratively ($k=1,...,K-1$) generate $P$ trial and admissible (i.e.,
check the ''\emph{Admissibility Condition}'') clusterings by applying
the integer-coded \emph{GA} operators \cite{Rocca 2020a}\cite{Rocca 2020b}.
For each $p$-th ($p=1,...,P$) tiling, set $\widehat{t}=\left(k-1\right)\times P+p$
and $\hat{\mathbf{c}}\leftarrow\mathbf{c}^{\left(\widehat{t}\right)}$
to determine $\hat{\bm{\alpha}}$ and $\hat{\bm{\beta}}$ as well
as $\widehat{\mathbf{w}}$ according to (\ref{eq:_sub-array.amplitudes-phases_EM})
and (\ref{eq:_hybrid_array}), respectively;%
\footnote{\noindent The number of tilings generated at the $i$-th ($i=1,...,I$)
\emph{O-DCTM} iteration is equal to $T_{O-DCTM}^{\left(i\right)}=P\times K$,
thus the total number of domino-clusterings processed by the \emph{O-DCTM}
turns out to be $T_{O-DCTM}=I\times P\times K$.%
}
\end{itemize}
\item \textbf{Step 2.4} - \emph{Optimal} $\mathcal{S}^{(i)}$\emph{-Tiling
Selection}. Define the $i$-th ($i=1,...,I$) optimal domino-tiling,
$\mathbf{c}^{best}$, as the one with the minimum value of the cost
function obtained so far ($\mathbf{c}^{best}$ $=$ $\arg$\{ $\min_{\widehat{t}=1,...,T^{(i)}}$
{[}$\Phi\left(\mathbf{c}^{\left(\widehat{t}\right)}\right)${]}\});
\item \textbf{Step 2.5} - \emph{Sub-Aperture Area Update.} Update the $(i+1)$-th
aperture area to be tiled, $\mathcal{R}^{(i+1)}\leftarrow\mathcal{R}^{(i)}$,
by removing the pixels covered by the optimal solution found at Step
2.4, $\mathbf{c}^{best}$ (Fig. 5). If the aperture $\mathcal{A}$
is fully tiled (i.e., $i=I$ and $\mathcal{R}^{\left(i+1\right)}=\emptyset$),
stop the iterative process and output the final array layout, $\mathbf{c}^{opt}=\mathbf{c}^{best}$.
Otherwise, increase the \emph{DCTM} iteration index ($i\leftarrow i+1$),
consider the new partition $\mathcal{S}^{(i)}$ to be clustered according
to the raster-scan scheme shown in Fig. 3 {[}e.g., Fig. 5 ($i=2$){]},
and goto Step 2.2.
\end{itemize}

\section{Numerical Assessment\label{sec:Numerical-Assessment}}

\noindent This section is devoted to assess the reliability of the
\emph{DCTM} as well as its effectiveness to handle large \emph{PA}s.
The first numerical example is then aimed at validating the \emph{DCTM}
by comparing its performance with that one of the \emph{ETM} \cite{Anselmi 2017}
when tiling small-size \emph{PA}s so that this latter method can be
executed in a feasible amount of time. More specifically, a square
($M=N$) array of $M\times N=8\times8$ equally-spaced ($d_{x}=d_{y}=\frac{\lambda}{2}$)
isotropic (i.e., $\mathbf{g}_{mn}\left(u,v\right)=\frac{1}{\sqrt{2}}\left(\widehat{\mathbf{u}}+\widehat{\mathbf{v}}\right)$;
$m=1,...,M$; $n=1,...,N$) elements has been considered. By choosing
the power pattern mask $\Psi\left(u,v\right)$ as in Fig. 6(\emph{a}),
the reference amplitude coefficients, $\bm{\alpha}^{ref}$, in Fig.
6(\emph{b}) have been computed with the convex programming (\emph{CP})
optimization strategy in \cite{Bucci 2002}. The phase terms have
been set to $0$ ($\bm{\beta}^{ref}=\mathbf{0}$) since the mask is
symmetric as well as broadside directed, thus real-valued reference
excitations are enough for affording the pattern $\mathcal{P}^{ref}\left(u,v\right)$
in Fig. 6(\emph{c}), which fulfils the radiation requirements at hand
{[}i.e., $\mathcal{P}^{ref}\left(u,v\right)\le\Psi\left(u,v\right)${]}.
The \emph{ETM} \cite{Anselmi 2017} has then been applied and the
whole set of $T=1.29\times10^{7}$ different domino-clustered configurations,
$\mathbf{C}$, has been generated by computing the corresponding cost
function values, \{$\Phi^{\left(t\right)}\triangleq\Phi\left(\mathbf{c}^{\left(t\right)}\right)$;
$t=1,...,T$\}, with (\ref{eq:_cost.function}), as well. Running
the \emph{ETM} on a computer equipped with an Intel 2.10 GHz Xeon
CPU and 64 Gb of RAM, the global optimal tiling, $\mathbf{c}_{ETM}^{opt}$
($\mathbf{c}_{ETM}^{opt}$ $=$ $\mathbf{c}^{min}$, $\mathbf{c}^{min}$
$\triangleq$ $\arg$\{$\min_{t=1,...,T}$ {[}$\Phi\left(\mathbf{c}^{\left(t\right)}\right)${]}\})
{[}Fig. 9(\emph{b}){]}, has been found in $\tau_{ETM}\approx22$ {[}days{]}.
Figure 7(\emph{a}) shows the behavior of $\Phi^{\left(t\right)}$
(\ref{eq:_cost.function}) versus the solution index sorted from that
of the worst tiling $\mathbf{c}^{max}$ ($\mathbf{c}^{max}$ $\triangleq$
$\arg$\{$\max_{t=1,...,T}$ {[}$\Phi\left(\mathbf{c}^{\left(t\right)}\right)${]}\})
to that of the best one $\mathbf{c}^{min}$, the minimum value of
the cost function $\Phi^{min}$ {[}$\Phi^{min}\triangleq\Phi\left(\mathbf{c}^{min}\right)${]}
at $\mathbf{c}_{ETM}^{opt}$ being equal to $\Phi_{ETM}^{opt}=1.14\times10^{-3}$.
Afterwards, the \emph{E-DCTM} has been used on the same benchmark
by setting the size of the partitions, \{$\mathcal{S}^{(i)}$; $i=1,...,I$\},
to $\widehat{M}\times\widehat{N}=2\times2$ such that $I=16$. With
reference to a non-optimized software implementation of the procedure
detailed in Sect. 2.2, the \emph{E-DCTM} ended in $\tau_{E-DCTM}\approx28$
{[}sec{]} after $T_{E-DCTM}=47$ ($T_{DCTM}\triangleq\sum_{i=1}^{I}T^{\left(i\right)}$)
evaluations of the cost function (\ref{eq:_cost.function}) by allowing
an impressive computational saving with respect to the \emph{ETM}
(i.e., $\frac{\tau_{ETM}}{\tau_{E-DCTM}}\approx68\times10^{3}$ and
$\frac{T_{ETM}}{T_{E-DCTM}}\approx25\times10^{3}$). More important,
the cost function value, $\Phi_{E-DCTM}^{opt}$, of the (\emph{E-DCTM})-optimized
layout, $\mathbf{c}_{E-DCTM}^{opt}$ {[}Fig. 9(\emph{c}){]}, turns
out to be almost coincident with the global optimum one {[}i.e., $\Phi_{E-DCTM}^{opt}=1.17\times10^{-3}$
vs. $\Phi_{ETM}^{opt}=1.14\times10^{-3}$ - Fig. 7(\emph{a}){]}, the
gap being very small (i.e., $\frac{\left|\Phi_{ETM}-\Phi_{E-DCTM}\right|}{\Phi_{E-DCTM}}\approx3$
\%). Such a slight deviation is pictorially pointed out by the plots
of the corresponding power patterns in Figs. 9(\emph{d})-9(\emph{e})
together with theirs cuts along the principal planes, $\phi=0$ {[}deg{]}
($v=0$) and $\phi=90$ {[}deg{]} ($u=0$), in Fig. 9(\emph{a}). For
the sake of completeness, the power pattern features, namely the sidelobe
level (\emph{SLL}), the peak directivity (\emph{D}), and the half-power
beam-width (\emph{HPBW}) along the principal planes for both the \emph{ETM}-tiled
arrays and the reference \emph{FP} one are reported in Tab. I. As
it can be inferred, the \emph{SLL} of the \emph{E-DCTM} pattern is
$7.05$ {[}dB{]} higher than that of the reference one, but it deviates
only $0.51$ {[}dB{]} from that of the \emph{ETM} solution (Tab. I).

\noindent As for the behavior of the \emph{E-DCTM} synthesis process,
it is interesting to observe in Fig. 7(\emph{b}) that the mask matching
metric (\ref{eq:_cost.function}) gets worse during the $I$ loops
since the single elementary radiators of the initial \emph{FP} array
layout are sequentially replaced by domino clusters as illustrated
in Fig. 8.

\noindent The second example is aimed at comparing the \emph{DCTM}
with other competitive state-of-the-art tiling techniques available
in the reference literature. Towards this end, the problem of domino-tiling
a rectangular array of $M\times N=22\times12$ elements, addressed
in \cite{Anselmi 2017}\cite{Yang 2021}, has been selected as benchmark.
In \cite{Anselmi 2017}\cite{Yang 2021}, the excitations of the \emph{FP}
reference array have been defined by considering separable distributions
and a Dolph-Chebyshev pattern \cite{Mailloux 2018}\cite{Elliot 2003}
with $SLL=-20.0$ {[}dB{]} to afford a power pattern, $\mathcal{P}^{ref}\left(u,v\right)$,
with main-lobe pointing towards broadside {[}i.e., $\left(u_{0},\, v_{0}\right)=\left(0,0\right)${]}.
For the sake of comparison, while taking into account the trade-off
relationship between \emph{SLL} and \emph{HPBW} \cite{Mailloux 2018}\cite{Haupt 2010},
the power mask $\Psi\left(u,v\right)$ has been tailored to force
the synthesis of a pattern with \emph{HPBW} values smaller or at most
equal to those of the reference solutions in \cite{Anselmi 2017}\cite{Yang 2021}
since - unlike (\ref{eq:_cost.function}) - the goal there was the
design of a domino-tiled layout with minimum \emph{SLL}. By keeping
the same $\mathcal{A}$-partitioning setup of the previous test case
(i.e., $\widehat{M}\times\widehat{N}=2\times2$), the layout outputted
at the convergence of the \emph{E-DCTM} synthesis is shown in Fig.
10(\emph{a}) as a color map of the amplitudes of the tiles, $\hat{\bm{\alpha}}_{E-DCTM}^{opt}$,
while the corresponding radiated power pattern in $\Omega$ and the
pattern cuts along the principal planes are given in Fig 10(\emph{b})
and Fig. 10(\emph{c}), respectively. It is worth noticing that the
\emph{SLL} of the (\emph{E-DCTM})-tiled array turns out to be very
close to that of the solutions in \cite{Anselmi 2017}\cite{Yang 2021}
(i.e., $SLL_{E-DCTM}-SLL_{[Anselmi\,2017]}$ $=$ $SLL_{E-DCTM}-SLL_{[Yang\,2021]}$
$=$ $0.22$ {[}dB{]} - Tab. II) despite the \emph{DCTM} synthesis
is not carried out at a time over the whole aperture $\mathcal{A}$.
Moreover, once again, the computational cost of the divide and conquer
strategy turns out to be lower ($\frac{\tau_{[Anselmi\,2017]}}{\tau_{E-DCTM}}\approx482$
and $\frac{\tau_{[Yang\,2021]}}{\tau_{E-DCTM}}\approx1.5$ - Tab.
II).

\noindent Open questions in using the \emph{DCTM} are the optimal
choice of the partition size, $\widehat{M}\times\widehat{N}$, as
well as that of the threshold $\eta_{th}$ for using either the \emph{E}-
or or the \emph{O}- version of the \emph{DCTM}. It is clear that these
parameters define the trade-off between the overall computational
burden and the effective sampling of the $T$-size solution space
(i.e., the possibility to reach the optimal admissible tiling of the
array aperture or domino layouts very close to it). On the other hand,
they are also strongly connected. In order to derive suitable guidelines
for the setup of such calibration parameters, the performance of both
the \emph{E-DCTM} and the \emph{O-DCTM} have been evaluated by choosing
a square ($M=N$) aperture $\mathcal{A}$ of $M\times N=24\times24$
$\frac{\lambda}{2}$-spaced elements partitioned into $I$ square
($\widehat{M}=\widehat{N}$) sub-areas, while varying $\eta$ subject
to the condition $\frac{\widehat{M}}{M}=\frac{\widehat{N}}{N}$ ($\to\eta=\frac{\widehat{M}}{M}=\frac{\widehat{N}}{N}$).
The power mask $\Psi\left(u,v\right)$ has been set as shown in Fig.
12 and the \emph{FP} reference array, whose radiation indexes are
reported in Tab. III, has been generated with the \emph{CP} method
\cite{Bucci 2002}. As for the \emph{O-DCTM}, the \emph{GA} control
parameters have been set according to \cite{Rocca 2009}: $P=3\times\widehat{M}\times\widehat{N}$,
$p_{c}=0.9$ ($p_{c}$ being the crossover probability), $p_{m}=0.01$
($p_{m}$ being the mutation probability), and $K=1000$ {[}$\to$
$T_{O-DCTM}^{\left(i\right)}=3000\times\widehat{M}\times\widehat{N}$
($i=1,...,I$){]}.

\noindent The results of such an analysis are summarized in Fig. 11(\emph{a})
where the behavior of $\Phi^{opt}$ and $T$ versus $\eta$ are reported.
By just observing the slope of $T_{E-DCTM}$ in Fig. 11(\emph{a}),
it is evident that the \emph{E-DCTM} can efficiently handle only very
small partitions of $\mathcal{A}$. For instance, the number of cost
function evaluations when $\eta=0.25$ ($\to$ $\widehat{M}\times\widehat{N}=6\times6$
and $I=3$) is very huge (i.e., $T_{E-DCTM}>8\times10^{5}$ - Tab.
III). Differently, the \emph{CPU}-time for the \emph{O-DCTM} synthesis
is feasible also for wider sub-apertures {[}i.e., $\eta>0.25$ - Fig.
11(\emph{a}){]} and the arising tilings (Fig. 14) optimally fulfill
the mask constraints ($\Phi_{O-DCTM}^{opt}\le6.0\times10^{-6}$ -
Fig. 12 and Tab. III). It is also worth pointing out that even when
$\eta\le0.25$, the \emph{O-DCTM} performs closely to the \emph{E-DCTM}
by synthesizing different tiled configurations (Fig. 13), but similar
in terms of radiated power patterns (Fig. 12 - Tab. III). Accordingly,
the switching threshold has been set to $\eta_{th}=0.25$.

\noindent As for the calibration of the partitions size, $\widehat{M}\times\widehat{N}$,
the following metric\begin{equation}
\chi_{\Delta}=\frac{1}{2}\left\{ \frac{\Phi_{\Delta}^{opt}}{\max_{\Delta}\left[\Phi_{\Delta}^{opt}\right]}+\frac{T_{\Delta}}{\max_{\Delta}\left[T_{\Delta}\right]}\right\} \label{eq:_sub-aperture.size.choice}\end{equation}
has been considered as an indicator function and its behaviour versus
the partition aspect-ratio $\Delta$ ($\Delta\triangleq\frac{\widehat{M}\times\widehat{N}}{M\times N}$)
has been analyzed. Figure 11(\emph{b}) shows that the minimum of $\chi_{\Delta}$
arises when $\Delta=\frac{1}{16}$, thus the value $\Delta^{opt}=\frac{1}{16}$
($\to$ $\left(\widehat{M}\times\widehat{N}\right)^{opt}=\frac{M\times N}{16}$)
has been chosen for the optimal sizing of the $I$ partitions, \{$\mathcal{S}^{(i)}$;
$i=1,...,I$\}, of the aperture $\mathcal{A}$.

\noindent The next experiments are concerned with the synthesis of
very large \emph{PA}s for high-data rate up-link communications between
Earth stations on mobile platforms (\emph{ESOMP}) and \emph{LEO} satellites.
These antenna systems require a value of the effective isotropic radiated
power (\emph{EIRP}) \cite{Mailloux 2018}, $E\left(\theta,\phi\right)$
( $E\left(u,v\right)\triangleq\Upsilon\frac{4\pi\mathcal{P}\left(u,v\right)}{\int_{\Omega}\frac{\mathcal{P}\left(u,v\right)}{\sqrt{1-u^{2}-v^{2}}}dudv}$,
$\Upsilon$ being the input power), when pointing towards broadside,
$\left(\theta_{0},\,\phi_{0}\right)=\left(0,0\right)$ {[}deg{]},
greater than $46.0$ {[}dBW{]} ($E_{dB}\left(\theta_{0},\phi_{0}\right)>46.0$
{[}dBW{]}, $E_{dB}\left(\theta_{0},\phi_{0}\right)\triangleq10\times\log\left[E_{dB}\left(\theta_{0},\phi_{0}\right)\right]$)
and the possibility to steer the beam up to $\theta^{max}=60$ {[}deg{]}
from broadside. Moreover, the \emph{EIRP} pattern must fit the ETSI
EN-303-978 mask \cite{ETSI}, $\Psi^{ETSI}\left(\theta,\,\phi\right)$,
in the whole scanning cone (i.e., $0\le\theta\le\theta^{max}$ and
$0\leq\phi<360$ {[}deg{]}). For illustrative purposes, the color
maps of $\Psi^{ETSI}\left(u,v\right)$ when pointing the main beam
towards broadside {[}i.e., $\left(u_{0},\, v_{0}\right)=\left(0,0\right)${]}
and along $\left(\theta_{0},\,\phi_{0}\right)=\left(60,0\right)$
{[}deg{]} {[}$\to$ $\left(u_{0},\, v_{0}\right)=\left(\frac{\sqrt{3}}{2},0\right)${]}
are shown in Fig. 15(\emph{a}) and Fig. 15(\emph{b}), respectively.

\noindent Given these requirements, the reference \emph{FP} array
has been defined to guarantee the absence of grating lobes within
the visible range, $\Omega$, whatever the main-lobe pointing direction,
and to generate an \emph{EIRP} pattern, $E\left(\theta,\phi\right)$,
compliant with the \emph{ETSI} masks, $\Psi^{ETSI}\left(\theta,\,\phi\right)$,
when steering up to the maximum scan angle ($0\le\theta\le\theta^{max}$
and $0\leq\phi<360$ {[}deg{]}). Accordingly, the \emph{FP} layout
turned out to be composed of $M\times N=80\times80$ elements spaced
by $d_{x}=d_{y}=0.52\lambda$ and uniformly fed (i.e., $\alpha_{mn}^{ref}=1.0$;
$m=1,...,M$; $n=1,...,N$). Moreover, the ($m,n$)-th ($m=1,...,M$;
$n=1,...,N$) phase reference value (\ref{eq:_sub-array.amplitudes-phases_EM})
has been set to\begin{equation}
\beta_{mn}^{ref}=-\frac{2\pi}{\lambda}\left(x_{mn}u_{0}+y_{mn}v_{0}\right).\label{eq:_reference.phase}\end{equation}
Furthermore, by assuming each ($m,n$)-th ($m=1,...,M$; $n=1,...,N$)
radiating element to have an element pattern equal to $\mathbf{g}_{mn}\left(u,v\right)=\sqrt[4]{\frac{1-u^{2}-v^{2}}{2}}\left(\mathbf{\hat{u}}+\mathbf{\hat{v}}\right)$,
the peak directivity of the reference antenna is equal to $D_{dB}=40.32$
{[}dBi{]} ($D_{dB}\triangleq10\times\log D$) at the maximum scan
(Tab. IV). Thus, since the \emph{EIRP} pattern is given by $E\left(\theta,\phi\right)=\Upsilon\times D\left(\theta,\phi\right)$,
the value of $\Upsilon$ has been set to $\Upsilon=4$ {[}W{]} (i.e.,
$\Upsilon_{dB}=6$ {[}dBW{]}) for fulfilling the \emph{ETSI} requirement
on the \emph{EIRP}.

\noindent Owing to the array size ($M\times N=6400$ elements), the
cardinality of the solution space composed by the whole set of $T$
admissible domino tilings of $\mathcal{A}$ is very huge ($T\ge1.0\times10^{200}$)
and an exhaustive search with either the \emph{ETM} or its divide-and-conquer
version (\emph{E-DCTM}) as well as the application of the \emph{OTM}
\cite{Anselmi 2017} would have been computationally unfeasible. Therefore,
the \emph{O-DCTM} has been used by setting $\Delta^{opt}=\frac{1}{16}$
so that the size of the partitions of the array aperture resulted
$\widehat{M}\times\widehat{N}=20\times20$. In less than 2 days (i.e.,
$\tau_{O-DCTM}\approx46$ {[}hours{]}), the irregular domino tiled
configuration $\mathbf{c}_{O-DCTM}^{opt}$ shown in Fig. 16(\emph{a})
has been synthesized. Such a layout generates the \emph{EIRP} patterns
in Figs. 16(\emph{b})-16(\emph{c}) towards broadside and those in
Figs. 17(\emph{c})-17(\emph{e}) and Figs. 17(\emph{d})-17(\emph{f})
when steering the beam at $\left(\theta_{0},\,\phi_{0}\right)=\left(60,0\right)$
{[}deg{]} and $\left(\theta_{0},\,\phi_{0}\right)=\left(60,90\right)$
{[}deg{]}, respectively. As for these two latter cases, the phase
excitations, $\bm{\beta}_{O-DCTM}^{opt}$, are those depicted in Fig.
17(\emph{a}) and Fig. 17(\emph{b}), respectively.

\noindent As it can be observed in Fig. 16(\emph{c}) and Figs. 17(\emph{e})-17(\emph{f}),
the \emph{O-DCTM} pattern, $E^{O-DCTM}\left(\theta,\phi\right)$,
is generally fully compliant with the \emph{ETSI} requirements, while
the few violations of the \emph{ETSI} mask, $\Psi^{ETSI}\left(u,\, v\right)$,
are almost negligible even along the most challenging scanning directions,
the pattern matching error $\Phi_{O-DCTM}^{opt}$ being kept below
$10^{-11}$ (i.e., $\Phi_{O-DCTM}^{opt}\approx7.5\times10^{-12}$
- Tab. IV). Moreover, as expected, the performance towards broadside,
$\left(u_{0},\, v_{0}\right)=\left(0,0\right)$ {[}Fig. 16(\emph{c}){]},
is ideal since there is no phase quantization.

\noindent Finally, it is worth pointing out that, analogously to all
previous experiments and in general, such an \emph{O-DCTM} domino-tiled
layout as well as all others \emph{DCTM}-synthesized array architectures,
reduces by half the number of \emph{TRM}s required by its \emph{FP}
reference counterpart (i.e., $Q=3200$ vs. $M\times N=6400$).

\section{Conclusions\label{sec:Conclusions}}

\noindent The problem of efficiently designing \emph{PA}s having antenna
elements organized into domino clusters and sub-array-only amplitude
and phase controls has been addressed by handling large apertures,
as well. Towards this end, an innovative domino-tiling method based
on a divide-and-conquer strategy has been proposed where the antenna
aperture has been subdivided into a set of partitions that have been
sequentially tiled, by properly exploiting customized techniques either
enumerative or optimization-based, until the complete clustering of
the whole array area.

\noindent From the numerical assessment, which also include comparisons
with state-of-the-art competitive tiling methods, the following main
outcomes can be drawn:

\begin{itemize}
\item thanks to the partitioning of the array aperture into sub-areas and
the use of soft-boundaries, the \emph{DCTM} is able to effectively
explore the solution space of the array tilings to find tiled layouts
close to the global optimum one with a non-negligible computational
saving with respect to other competitive state-of-the-art approaches;
\item the \emph{E-DCTM} turns out to be optimal (i.e., it finds the optimal
tiled-layout) and computationally-efficient with respect to both the
\emph{ETM} and the \emph{OTM} when dealing with small/medium arrays
and the condition $\eta_{th}<0.25$ holds true, while it is unfeasible
otherwise;
\item \noindent thanks to the sequential optimization-driven domino-clustering
of the partitions of the array aperture, the \emph{O-DCTM} turns out
to be a key enabling and reliable method for tiling large \emph{PA}s.
\end{itemize}
Future research activities, beyond the scope of this work, will investigate
the use of different tile shapes as well as the extension of the \emph{DCTM}
to other large aperture geometries of interest in relevant \emph{PA}
applications by also including additional manufacturing constraints
to simplify the beam-forming network implementation and/or to comply
with mechanical features of the installation site (e.g., blockage
effects).

\section*{Acknowledgment}

\noindent This work benefited from the networking activities carried
out within \textbf{}the Project DICAM-EXC (Grant L232/2016) funded
by the Italian Ministry of Education, Universities and Research (MUR)
within the 'Departments of Excellence 2023-2027' Program (CUP: E63C22003880001),
\textbf{}the Project SEME@TN - Smart ElectroMagnetic Environment in
TrentiNo funded by the Autonomous Province of Trento (CUP: C63C22000720003),
the Project AURORA - Smart Materials for Ubiquitous Energy Harvesting,
Storage, and Delivery in Next Generation Sustainable Environments
funded by the Italian Ministry for Universities and Research within
the PRIN-PNRR 2022 Program (CUP: E53D23014760001), the Project National
Centre for HPC, Big Data and Quantum Computing (CN HPC) funded by
the European Union - NextGenerationEU within the PNRR Program (CUP:
E63C22000970007), the Project Telecommunications of the Future {[}PE00000001
- program RESTART, Project 6GWINET (CUP: D43C22003080001), Project
MOSS (CUP: J33C22002880001), Project IN (CUP: J33C22002880001), Project
EMS-MMDV (CUP: J33C22002880001), Project TRIBOLETTO (CUP: B53C22003970001),
and Project SMART (CUP: E63C22002040007){]}, funded by European Union
under the Italian National Recovery and Resilience Plan (NRRP) of
NextGenerationEU, and the support of the Natural Science Basic Research
Program of Shaanxi Province (Grants No. 2022-JC-33, No. 2023-GHZD-35,
and No. 2024JC-ZDXM-25). Views and opinions expressed are however
those of the author(s) only and do not necessarily reflect those of
the European Union or the European Research Council. Neither the European
Union nor the granting authority can be held responsible for them.
A. Massa wishes to thank E. Vico and L. Massa for their never-ending
inspiration, support, guidance, and help.

~

\emph{This work has been submitted to the IEEE for possible publication.
Copyright may be transferred without notice, after which this version
may no longer be accessible}

\newpage
\section*{FIGURE CAPTIONS}

\begin{itemize}
\item \textbf{Figure 1.} Sketch of a domino-tiled \emph{PA} layout.
\item \textbf{Figure 2.} \emph{Illustrative Example} ($M=N=12$) - Sketch
of (\emph{a}) the checkerboard pattern of the array aperture $\mathcal{A}$
together with the external/internal vertices and the edges and (\emph{b})
the \emph{minimal} (domino) \emph{tiling} of $\mathcal{A}$ including
the values of the \emph{HF} on the vertices.
\item \textbf{Figure 3.} \emph{Illustrative Example} ($M=N=12$; $I=9$)
- Sketch of the partition of the array aperture $\mathcal{A}$ into
$I$ sub-areas, \{$\mathcal{S}^{(i)}$; $i=1,...,I$\}.
\item \textbf{Figure 4.} \emph{Illustrative Example} ($M=N=12$; $i=1$)
- Sketches of the partition of $\mathcal{A}$ under \emph{DCTM} domino-tiling,
$\mathcal{S}^{(i)}$, the area to be tiled, $\mathcal{R}^{(i)}$,
and the remaining area, $\mathcal{D}^{(i)}$ ($\mathcal{D}^{(i)}\triangleq\mathcal{R}^{(i)}-\mathcal{S}^{(i)}$).
\item \textbf{Figure 5.} \emph{Illustrative Example} ($M=N=12$; $i=2$)
- Sketches of the partition of $\mathcal{A}$ under \emph{DCTM} domino-tiling,
$\mathcal{S}^{(i)}$, the area to be tiled, $\mathcal{R}^{(i)}$,
and the remaining area, $\mathcal{D}^{(i)}$ ($\mathcal{D}^{(i)}\triangleq\mathcal{R}^{(i)}-\mathcal{S}^{(i)}$).
\item \textbf{Figure 6.} \emph{Numerical Validation} ($M=N=8$, Isotropic
Elements, $d_{x}=d_{y}=\frac{\lambda}{2}$, $\left(\theta_{0},\phi_{0}\right)=\left(0,0\right)$
{[}deg{]} {[}$\to$ $\left(u_{0},v_{0}\right)=\left(0,0\right)${]})
- Plots of (\emph{a}) the power pattern mask, $\Psi\left(u,v\right)$,
(\emph{b}) the amplitude distribution of the reference excitations,
$\bm{\alpha}^{ref}$, and (\emph{c}) the reference power pattern,
$\mathcal{P}^{ref}\left(u,v\right)$, in the ($u,v$)-domain $\Omega$.
\item \textbf{Figure 7.} \emph{Numerical Validation} ($M=N=8$, Isotropic
Elements, $d_{x}=d_{y}=\frac{\lambda}{2}$, $\left(\theta_{0},\phi_{0}\right)=\left(0,0\right)$
{[}deg{]}) - Behavior of the value of the pattern matching function,
$\Phi$, versus (\emph{a}) the {}``sorted'' solution index $t$
($1\le t\le T$) and (\emph{b}) the \emph{E-DCTM} iteration index
$i$ ($i=1,...,I$; $I=16$).
\item \textbf{Figure 8.} \emph{Numerical Validation} ($M=N=8$, Isotropic
Elements, $d_{x}=d_{y}=\frac{\lambda}{2}$, $\left(\theta_{0},\phi_{0}\right)=\left(0,0\right)$
{[}deg{]}; \emph{E-DCTM} - $\widehat{M}\times\widehat{N}=2\times2$
$\to$ $I=16$) - Sketch of $\mathbf{c}^{best}$ at the $i$-th \emph{E-DCTM}
iteration: (\emph{a}) $i=1$, (\emph{b}) $i=3$, (\emph{c}) $i=8$,
and (\emph{d}) $i=14$.
\item \textbf{Figure 9.} \emph{Numerical Validation} ($M=N=8$, Isotropic
Elements, $d_{x}=d_{y}=\frac{\lambda}{2}$, $\left(\theta_{0},\phi_{0}\right)=\left(0,0\right)$
{[}deg{]} {[}$\to$ $\left(u_{0},v_{0}\right)=\left(0,0\right)${]})
- Plots of (\emph{a}) the normalized power pattern, $\mathcal{P}\left(u,v\right)$,
along the principal planes $v=0$ (i.e., $\phi=0$ {[}deg{]}) and
$u=0$ (i.e., $\phi=90$ {[}deg{]}), (\emph{b})(\emph{c}) the amplitude
distribution of the clustered excitations, $\bm{\alpha}$, and (\emph{d})(\emph{e})
$\mathcal{P}\left(u,v\right)$ in the ($u,v$)-domain $\Omega$ synthesized
with (\emph{b})(\emph{d}) the \emph{ETM} and (\emph{c})(\emph{e})
the \emph{E-DCTM} ($\widehat{M}\times\widehat{N}=2\times2$ $\to$
$I=16$).
\item \textbf{Figure 10.} \emph{Comparative Assessment} ($M=22$, $N=12$,
Isotropic Elements, $d_{x}=d_{y}=\frac{\lambda}{2}$, $\left(\theta_{0},\phi_{0}\right)=\left(0,0\right)$
{[}deg{]} {[}$\to$ $\left(u_{0},v_{0}\right)=\left(0,0\right)${]};
\emph{E-DCTM} - $\widehat{M}\times\widehat{N}=2\times2$ $\to$ $I=66$)
- Plots of (\emph{a}) the amplitude distribution of the clustered
excitations, $\bm{\alpha}_{E-DCTM}^{opt}$, and the corresponding
normalized power pattern, $\mathcal{P}^{E-DCTM}\left(u,v\right)$,
(\emph{b}) in the ($u,v$)-domain $\Omega$ and (\emph{c}) along the
principal planes $v=0$ (i.e., $\phi=0$ {[}deg{]}) and $u=0$ (i.e.,
$\phi=90$ {[}deg{]}).
\item \textbf{Figure 11.} \emph{Numerical Assessment} ($M=N=24$, Isotropic
Elements, $d_{x}=d_{y}=\frac{\lambda}{2}$, $\left(\theta_{0},\phi_{0}\right)=\left(0,0\right)$
{[}deg{]}) - Behavior of (\emph{a}) the optimal value of the pattern
matching function, $\Phi^{opt}$, and the number of cost function
evaluations, $T_{DCTM}$, versus $\eta$ and (\emph{b}) the indicator
function $\chi_{\Delta}$ versus the partition aspect ratio $\Delta$.
\item \textbf{Figure 12.} \emph{Numerical Assessment} ($M=N=24$, Isotropic
Elements, $d_{x}=d_{y}=\frac{\lambda}{2}$, $\left(\theta_{0},\phi_{0}\right)=\left(0,0\right)$
{[}deg{]} {[}$\to$ $\left(u_{0},v_{0}\right)=\left(0,0\right)${]})
- Plots of the normalized power pattern $\mathcal{P}\left(u,v\right)$
along (\emph{a}) the $v=0$ (i.e., $\phi=0$ {[}deg{]}) and (\emph{b})
the $u=0$ (i.e., $\phi=90$ {[}deg{]}) planes.
\item \textbf{Figure 13.} \emph{Numerical Assessment} ($M=N=24$, Isotropic
Elements, $d_{x}=d_{y}=\frac{\lambda}{2}$, $\left(\theta_{0},\phi_{0}\right)=\left(0,0\right)$
{[}deg{]}) - Sketch of the domino tiling $\mathbf{c}^{opt}$ synthesized
with (\emph{a})(\emph{b})(\emph{c}) \emph{}the \emph{E-DCTM} and (\emph{d})(\emph{e})(\emph{f})
the \emph{O-DCTM} when (\emph{a})(\emph{d}) $\eta=\frac{1}{12}$,
(\emph{b})(e) $\eta=\frac{1}{6}$, and (\emph{c})(\emph{f}) $\eta=\frac{1}{4}$.
\item \textbf{Figure 14.} \emph{Numerical Assessment} ($M=N=24$, Isotropic
Elements, $d_{x}=d_{y}=\frac{\lambda}{2}$, $\left(\theta_{0},\phi_{0}\right)=\left(0,0\right)$
{[}deg{]}; \emph{O-DCTM}) - Sketch of the domino tiling $\mathbf{c}_{O-DCTM}^{opt}$
when (\emph{a}) $\eta=\frac{1}{3}$, (\emph{b}) $\eta=\frac{1}{2}$,
and (\emph{c}) $\eta=1$.
\item \textbf{Figure 15.} \emph{Numerical Assessment} ($M=N=80$, Directive
Elements, $d_{x}=d_{y}=0.52\lambda$) - \emph{ETSI} mask, $\Psi^{ETSI}\left(u,v\right)$,
when the main-lobe is pointed at (\emph{a}) broadside (i.e., $\left(u_{0},v_{0}\right)=\left(0,0\right)$
$\to$ $\left(\theta_{0},\phi_{0}\right)=\left(0,0\right)$ {[}deg{]})
and (\emph{b}) $\left(u_{0},v_{0}\right)=\left(\frac{\sqrt{3}}{2},0\right)$
($\to$ $\left(\theta_{0},\phi_{0}\right)=\left(60,0\right)$ {[}deg{]}).
\item \textbf{Figure 16.} \emph{Numerical Assessment} ($M=N=80$, Directive
Elements, $d_{x}=d_{y}=0.52\lambda$, $\left(\theta_{0},\phi_{0}\right)=\left(0,0\right)$
{[}deg{]} $\to$ $\left(u_{0},v_{0}\right)=\left(0,0\right)$; \emph{O-DCTM}
- $\widehat{M}\times\widehat{N}=20\times20$ $\to$ $I=16$) - Pictures
of (\emph{a}) the sketch of the domino tiling $\mathbf{c}_{O-DCTM}^{opt}$
and the corresponding \emph{EIRP} pattern, $E^{O-DCTM}\left(u,v\right)$,
(\emph{b}) in the ($u,v$)-domain $\Omega$ and (\emph{c}) along the
principal planes $v=0$ (i.e., $\phi=0$ {[}deg{]}) and $u=0$ (i.e.,
$\phi=90$ {[}deg{]}).
\item \textbf{Figure 17.} \emph{Numerical Assessment} ($M=N=80$, Directive
Elements, $d_{x}=d_{y}=0.52\lambda$; \emph{O-DCTM} - $\widehat{M}\times\widehat{N}=20\times20$
$\to$ $I=16$) - Plots of (\emph{a})(\emph{b}) the phase distribution
of the clustered excitations, $\bm{\beta}_{O-DCTM}^{opt}$, and the
corresponding \emph{EIRP} pattern , $E^{O-DCTM}\left(u,v\right)$,
(\emph{c})(\emph{d}) in the ($u,v$)-domain $\Omega$ and (\emph{e})(\emph{f})
along the principal planes $v=0$ (i.e., $\phi=0$ {[}deg{]}) and
$u=0$ (i.e., $\phi=90$ {[}deg{]}) when pointing the beam towards
(\emph{a})(\emph{c})(\emph{e}) $\left(u_{0},v_{0}\right)=\left(\frac{\sqrt{3}}{2},0\right)$
($\to$ $\left(\theta_{0},\phi_{0}\right)=\left(60,0\right)$ {[}deg{]})
and (\emph{b})(\emph{d})(\emph{f}) $\left(u_{0},v_{0}\right)=\left(0,\frac{\sqrt{3}}{2}\right)$
($\to$ $\left(\theta_{0},\phi_{0}\right)=\left(60,90\right)$ {[}deg{]}).
\end{itemize}

\section*{TABLE CAPTIONS}

\begin{itemize}
\item \textbf{Table I.} \emph{Numerical Validation} ($M=N=8$, Isotropic
Elements, $d_{x}=d_{y}=\frac{\lambda}{2}$, $\left(\theta_{0},\phi_{0}\right)=\left(0,0\right)$
{[}deg{]}) - Performance indexes.
\item \textbf{Table II.} \emph{Comparative Assessment} ($M=22$, $N=12$,
Isotropic Elements, $d_{x}=d_{y}=\frac{\lambda}{2}$, $\left(\theta_{0},\phi_{0}\right)=\left(0,0\right)$
{[}deg{]}) - Performance indexes.
\item \textbf{Table III.} \emph{Numerical Assessment} ($M=N=24$, Isotropic
Elements, $d_{x}=d_{y}=\frac{\lambda}{2}$, $\left(\theta_{0},\phi_{0}\right)=\left(0,0\right)$
{[}deg{]}) - Performance indexes.
\item \textbf{Table IV.} \emph{Numerical Assessment} ($M=N=80$, Directive
Elements, $d_{x}=d_{y}=0.52\lambda$) - Performance indexes.
\end{itemize}
\newpage
\begin{center}~\vfill\end{center}

\begin{center}\begin{tabular}{c}
\includegraphics[%
  width=0.80\columnwidth,
  keepaspectratio]{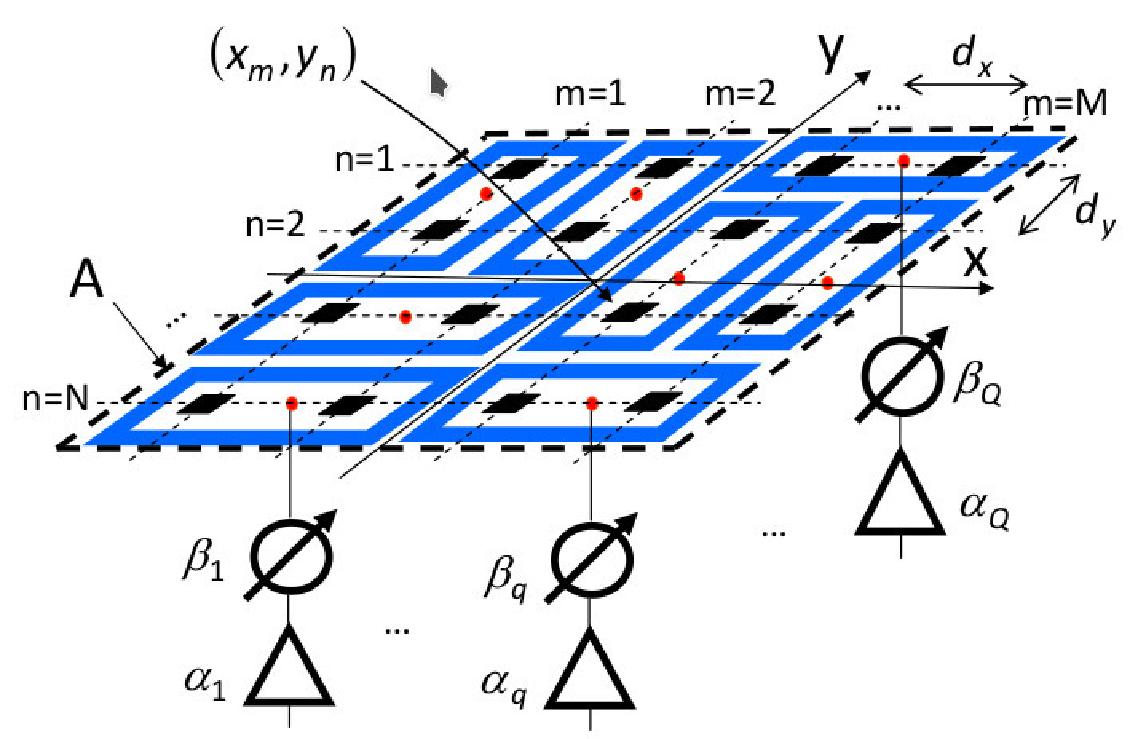}\tabularnewline
\end{tabular}\end{center}

\begin{center}~\vfill\end{center}

\begin{center}\textbf{Fig. 1 - N. Anselmi} \textbf{\emph{et al.}}\textbf{,}
\textbf{\emph{{}``}}A Divide-and-Conquer Tiling Method ...''\end{center}

\newpage
\begin{center}~\vfill\end{center}

\begin{center}\begin{tabular}{c}
\includegraphics[%
  width=0.60\columnwidth,
  keepaspectratio]{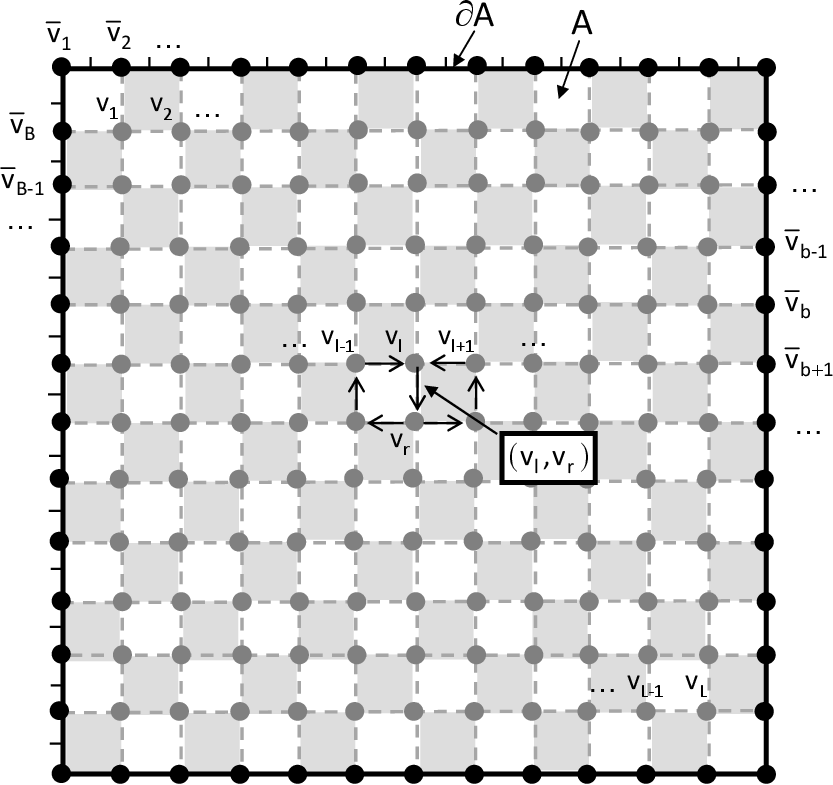}\tabularnewline
(\emph{a})\tabularnewline
\includegraphics[%
  width=0.60\columnwidth,
  keepaspectratio]{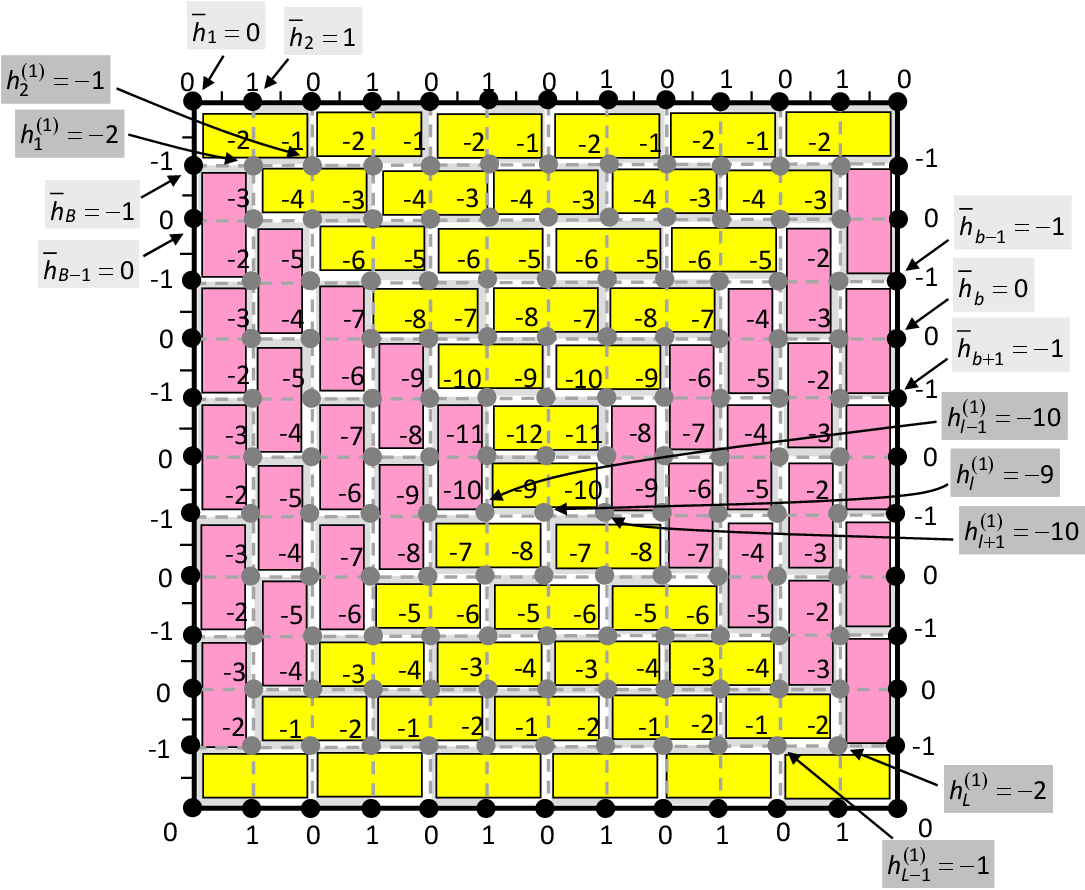}\tabularnewline
(\emph{b})\tabularnewline
\end{tabular}\end{center}

\begin{center}~\vfill\end{center}

\begin{center}\textbf{Fig. 2 - N. Anselmi} \textbf{\emph{et al.}}\textbf{,}
\textbf{\emph{{}``}}A Divide-and-Conquer Tiling Method ...''\end{center}

\newpage
\begin{center}~\vfill\end{center}

\begin{center}\begin{tabular}{c}
\includegraphics[%
  width=0.60\columnwidth,
  keepaspectratio]{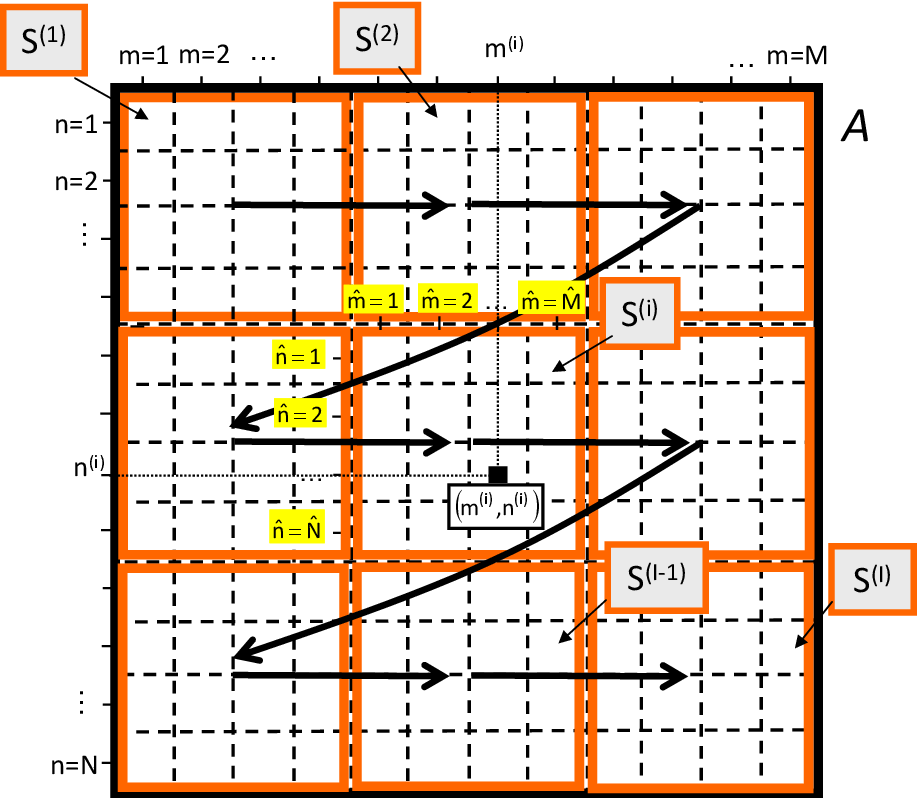}\tabularnewline
\end{tabular}\end{center}

\begin{center}~\vfill\end{center}

\begin{center}\textbf{Fig. 3 - N. Anselmi} \textbf{\emph{et al.}}\textbf{,}
\textbf{\emph{{}``}}A Divide-and-Conquer Tiling Method ...''\end{center}

\newpage
\begin{center}~\vfill\end{center}

\begin{center}\begin{tabular}{c}
\includegraphics[%
  width=0.70\columnwidth,
  keepaspectratio]{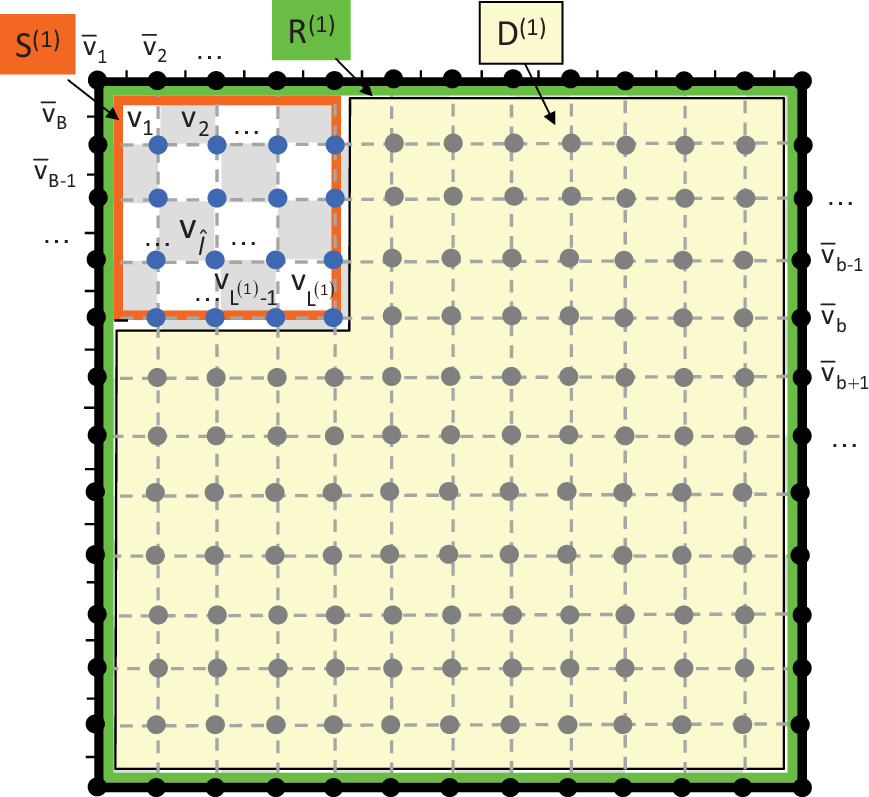}\tabularnewline
\end{tabular}\end{center}

\begin{center}~\vfill\end{center}

\begin{center}\textbf{Fig. 4 - N. Anselmi} \textbf{\emph{et al.}}\textbf{,}
\textbf{\emph{{}``}}A Divide-and-Conquer Tiling Method ...''\end{center}

\newpage
\begin{center}~\vfill\end{center}

\begin{center}\begin{tabular}{c}
\includegraphics[%
  width=0.65\columnwidth,
  keepaspectratio]{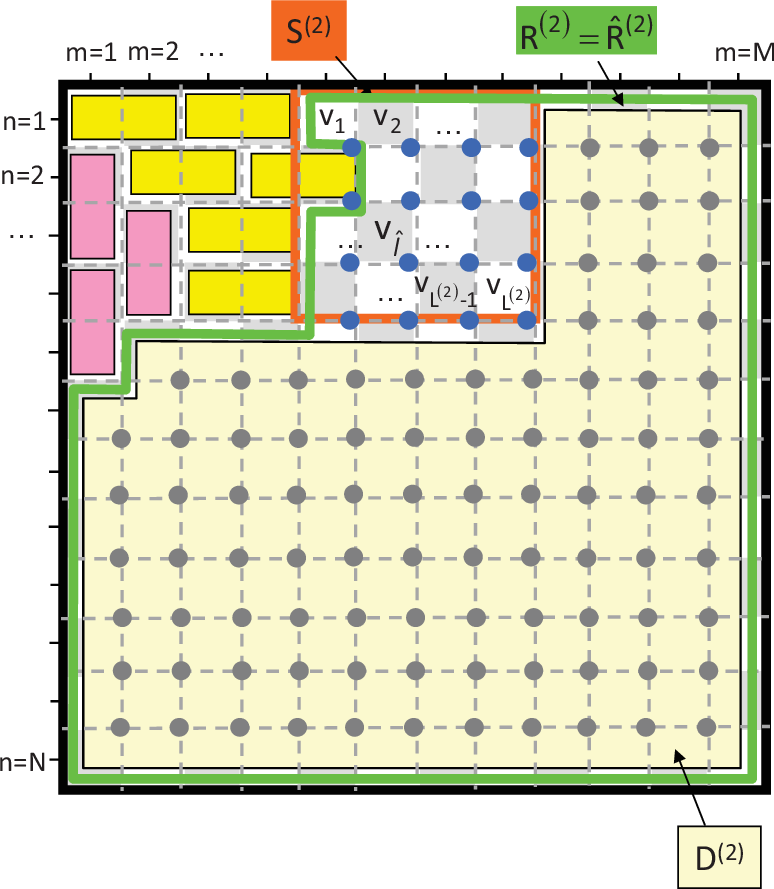}\tabularnewline
\end{tabular}\end{center}

\begin{center}~\vfill\end{center}

\begin{center}\textbf{Fig. 5 - N. Anselmi} \textbf{\emph{et al.}}\textbf{,}
\textbf{\emph{{}``}}A Divide-and-Conquer Tiling Method ...''\end{center}

\newpage
\begin{center}~\vfill\end{center}

\begin{center}\begin{tabular}{cc}
\multicolumn{2}{c}{ \includegraphics[%
  width=0.50\columnwidth,
  keepaspectratio]{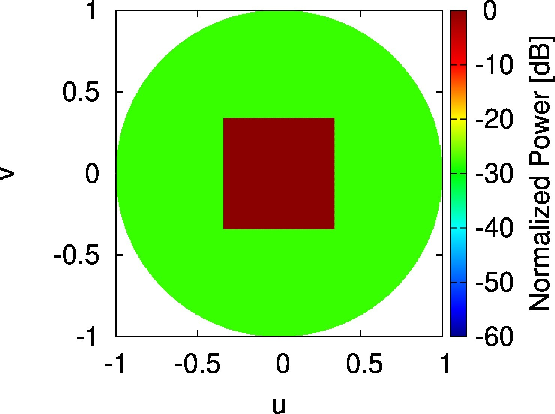}}\tabularnewline
\multicolumn{2}{c}{(\emph{a})}\tabularnewline
 \includegraphics[%
  width=0.50\columnwidth,
  keepaspectratio]{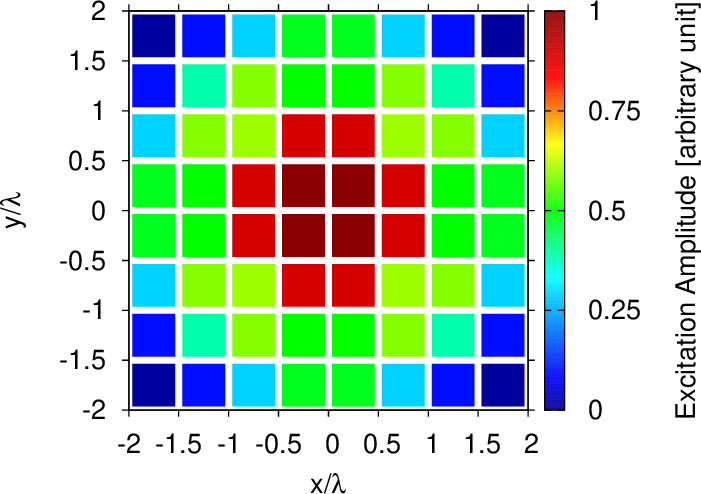}&
 \includegraphics[%
  width=0.50\columnwidth,
  keepaspectratio]{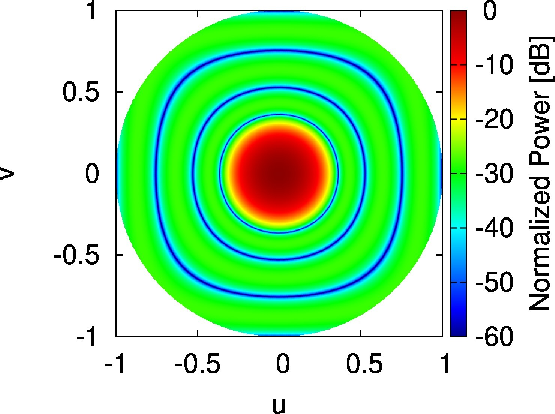}\tabularnewline
(\emph{b})&
(\emph{c})\tabularnewline
\end{tabular}\end{center}

\begin{center}~\vfill\end{center}

\begin{center}\textbf{Fig. 6 - N. Anselmi} \textbf{\emph{et al.}}\textbf{,}
\textbf{\emph{{}``}}A Divide-and-Conquer Tiling Method ...''\end{center}

\newpage
\begin{center}~\vfill\end{center}

\begin{center}\begin{tabular}{c}
\includegraphics[%
  width=0.80\columnwidth,
  keepaspectratio]{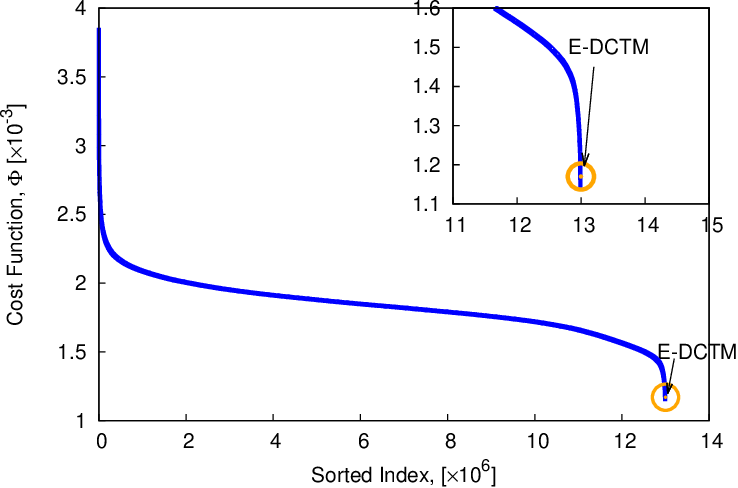}\tabularnewline
(\emph{a})\tabularnewline
\includegraphics[%
  width=0.80\columnwidth,
  keepaspectratio]{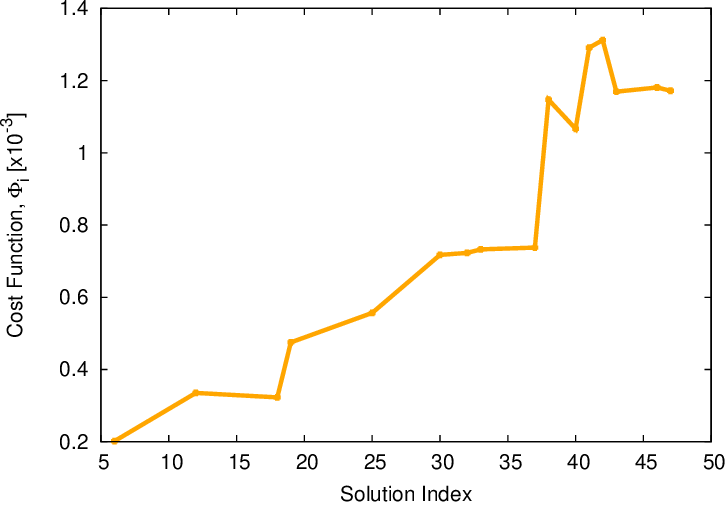}\tabularnewline
(\emph{b})\tabularnewline
\end{tabular}\end{center}

\begin{center}~\vfill\end{center}

\begin{center}\textbf{Fig. 7 - N. Anselmi} \textbf{\emph{et al.}}\textbf{,}
\textbf{\emph{{}``}}A Divide-and-Conquer Tiling Method ...''\end{center}

\newpage
\begin{center}~\vfill\end{center}

\begin{center}\begin{tabular}{cc}
\includegraphics[%
  width=0.40\columnwidth]{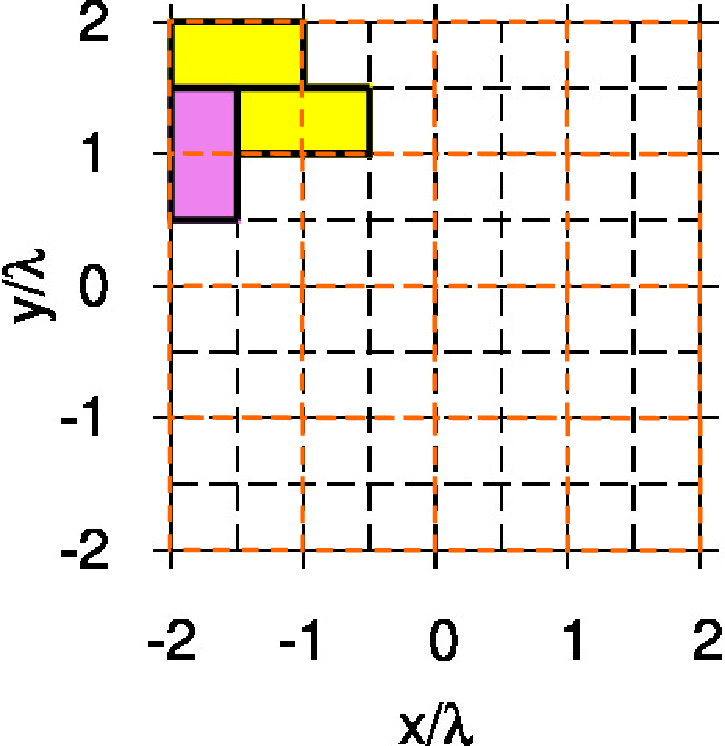}&
\includegraphics[%
  width=0.40\columnwidth]{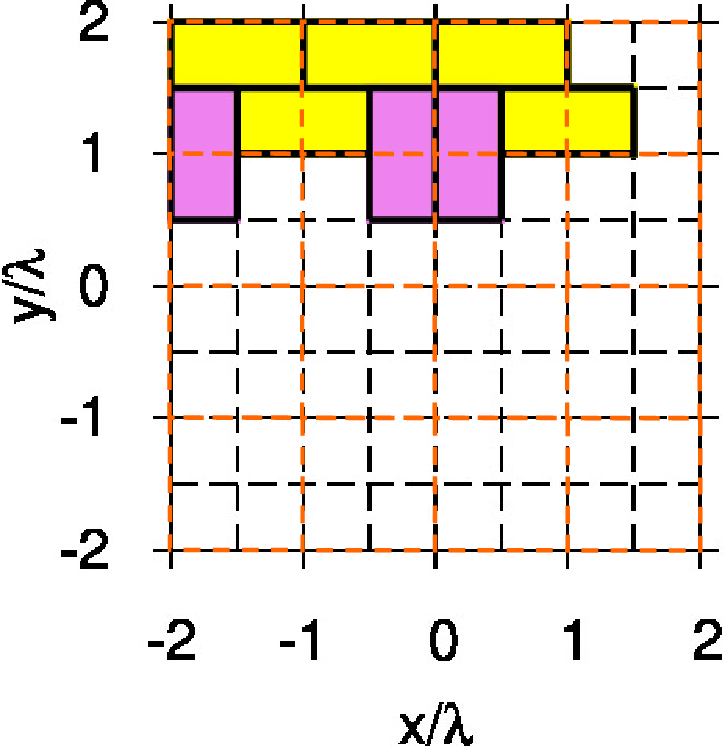}\tabularnewline
(\emph{a})&
(\emph{b})\tabularnewline
&
\tabularnewline
\includegraphics[%
  width=0.40\columnwidth]{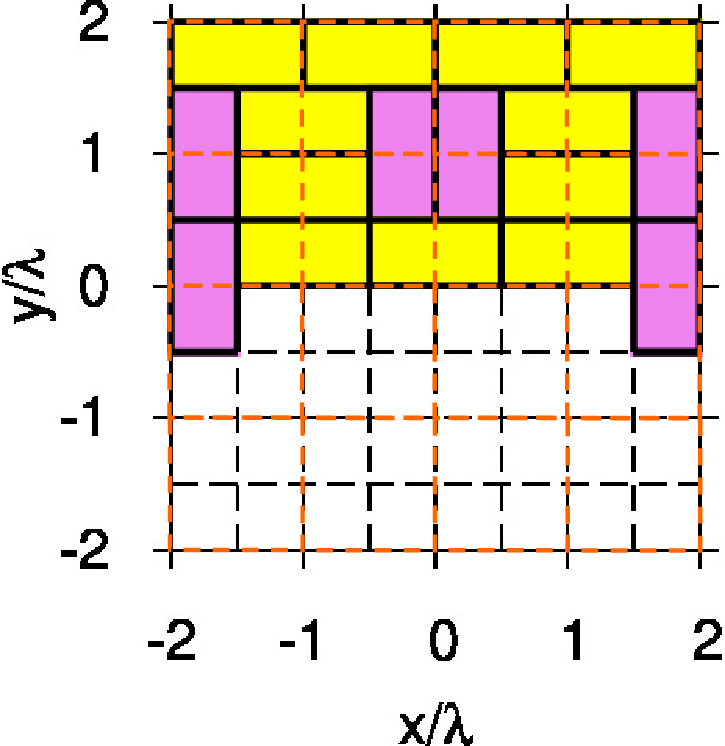}&
\includegraphics[%
  width=0.40\columnwidth]{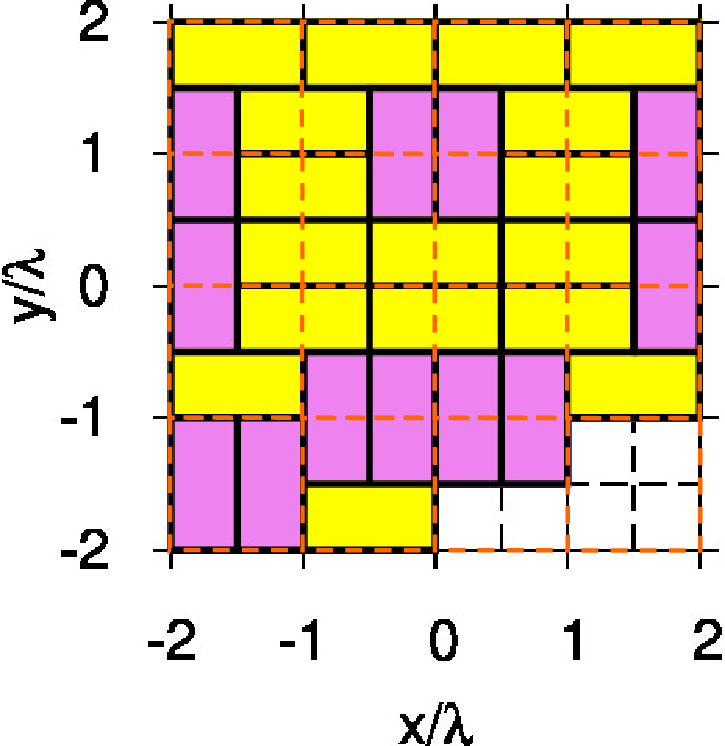}\tabularnewline
(\emph{c})&
(\emph{d})\tabularnewline
\end{tabular}\end{center}

\begin{center}~\vfill\end{center}

\begin{center}\textbf{Fig. 8 - N. Anselmi} \textbf{\emph{et al.}}\textbf{,}
\textbf{\emph{{}``}}A Divide-and-Conquer Tiling Method ...''\end{center}

\newpage
\begin{center}~\vfill\end{center}

\begin{center}\begin{tabular}{cc}
\multicolumn{2}{c}{ \includegraphics[%
  width=0.50\columnwidth,
  keepaspectratio]{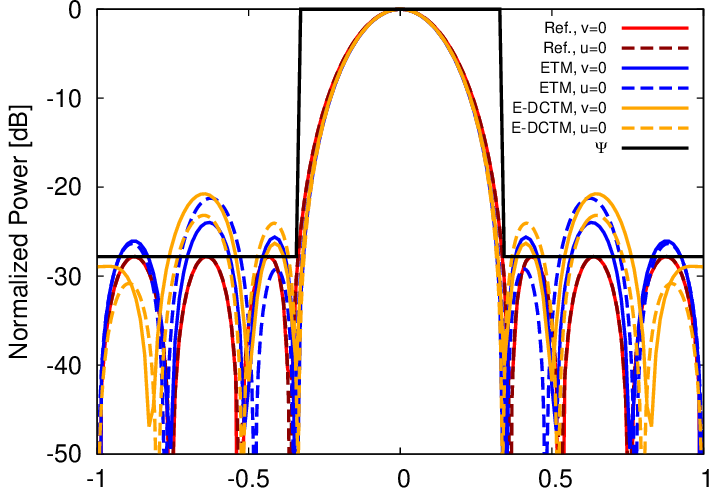}}\tabularnewline
\multicolumn{2}{c}{(\emph{a})}\tabularnewline
\includegraphics[%
  width=0.40\columnwidth,
  keepaspectratio]{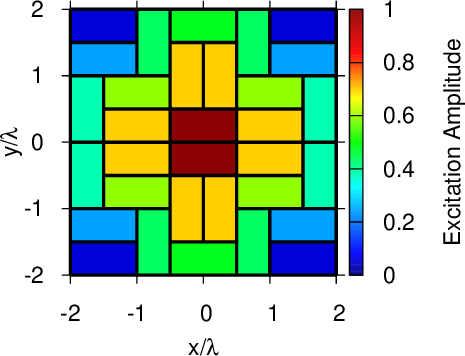}&
\includegraphics[%
  width=0.40\columnwidth,
  keepaspectratio]{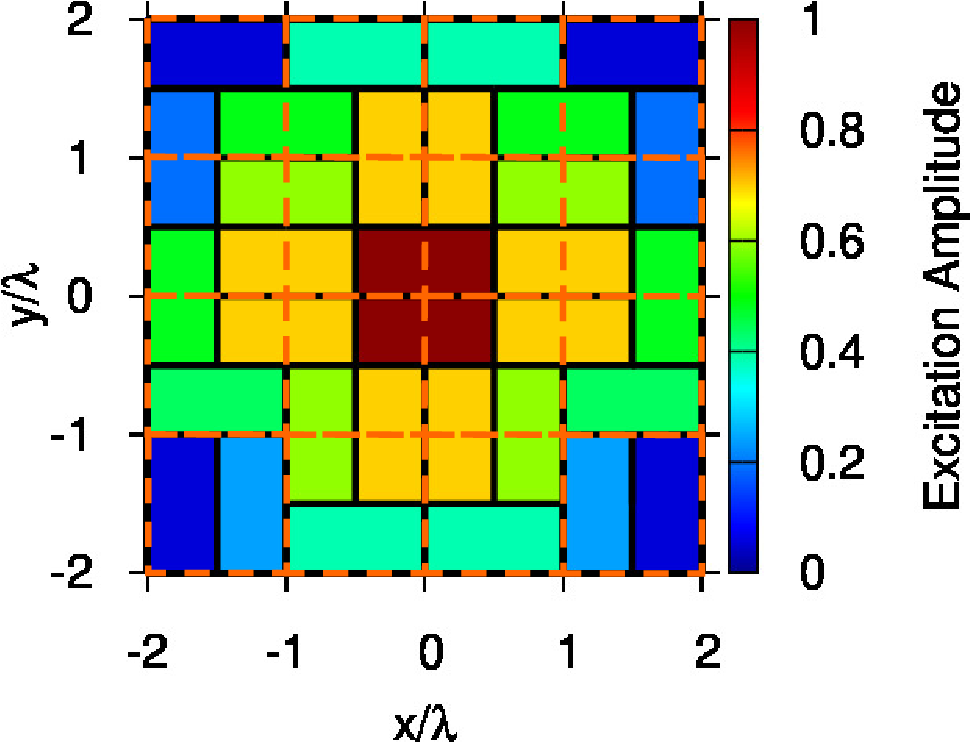}\tabularnewline
(\emph{b})&
(\emph{c})\tabularnewline
\includegraphics[%
  width=0.45\columnwidth,
  keepaspectratio]{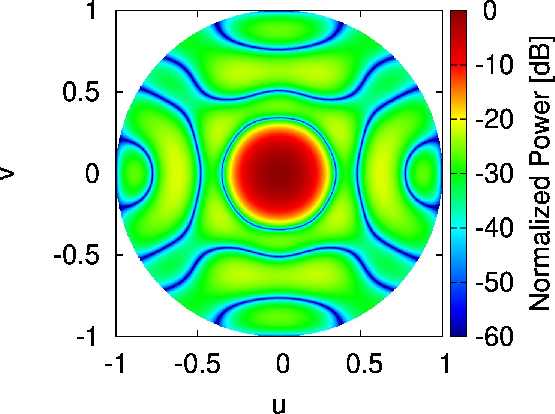}&
\includegraphics[%
  width=0.45\columnwidth,
  keepaspectratio]{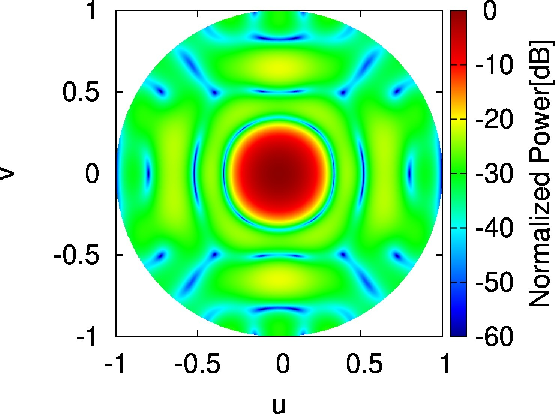}\tabularnewline
(\emph{d})&
(\emph{e})\tabularnewline
\end{tabular}\end{center}

\begin{center}~\vfill\end{center}

\begin{center}\textbf{Fig. 9 - N. Anselmi} \textbf{\emph{et al.}}\textbf{,}
\textbf{\emph{{}``}}A Divide-and-Conquer Tiling Method ...''\end{center}

\newpage
\begin{center}~\vfill\end{center}

\begin{center}\begin{tabular}{cc}
\multicolumn{2}{c}{\includegraphics[%
  width=0.70\columnwidth]{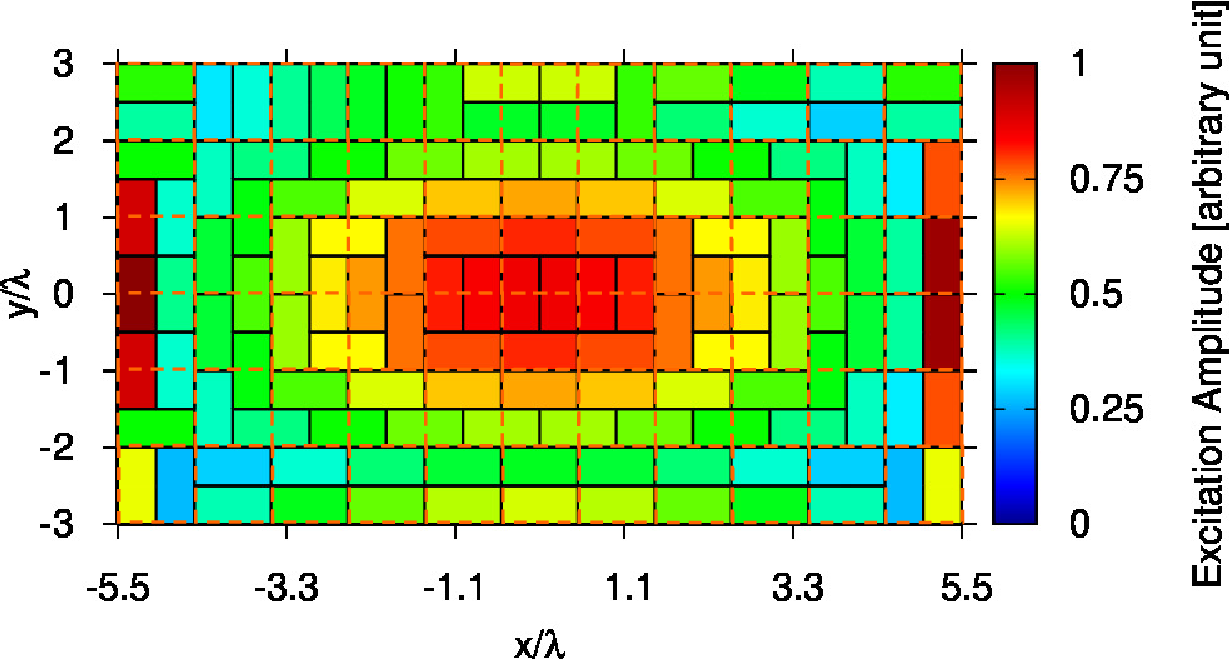}}\tabularnewline
\multicolumn{2}{c}{(\emph{a})}\tabularnewline
\includegraphics[%
  width=0.48\columnwidth]{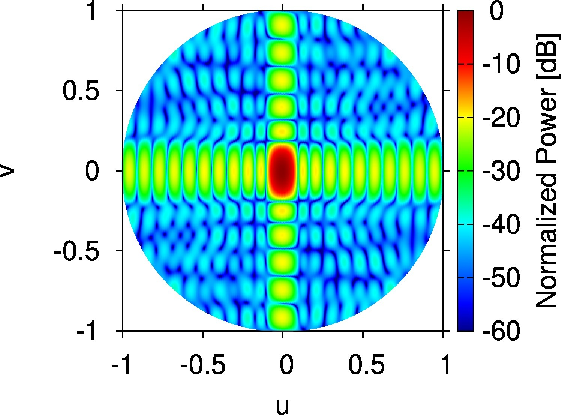}&
\includegraphics[%
  width=0.60\columnwidth]{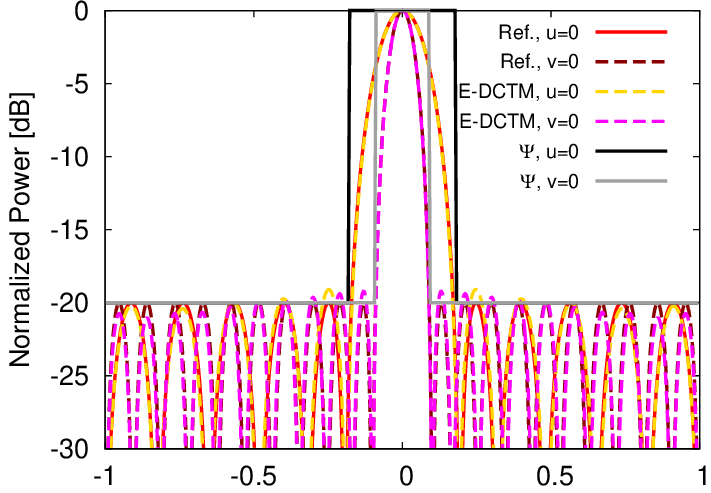}\tabularnewline
(\emph{b})&
(\emph{c})\tabularnewline
\end{tabular}\end{center}

\begin{center}~\vfill\end{center}

\begin{center}\textbf{Fig. 10 - N. Anselmi} \textbf{\emph{et al.}}\textbf{,}
\textbf{\emph{{}``}}A Divide-and-Conquer Tiling Method ...''\end{center}

\newpage
\begin{center}~\vfill\end{center}

\begin{center}\begin{tabular}{c}
\includegraphics[%
  width=0.60\columnwidth]{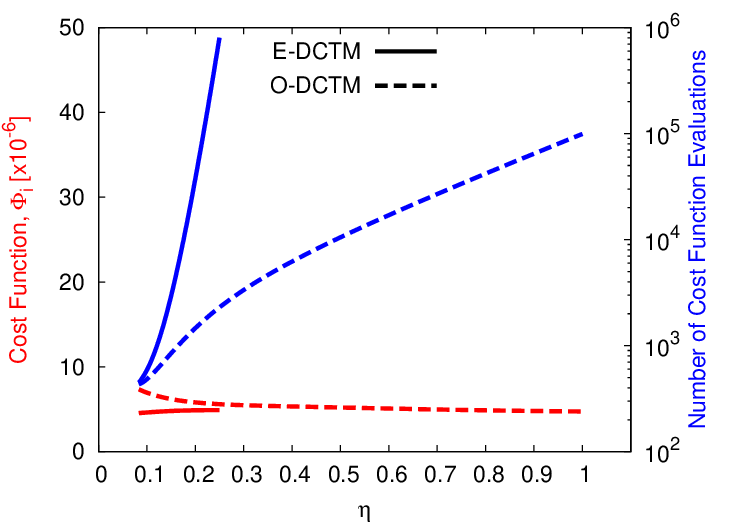}\tabularnewline
(\emph{a})\tabularnewline
\includegraphics[%
  width=0.60\columnwidth]{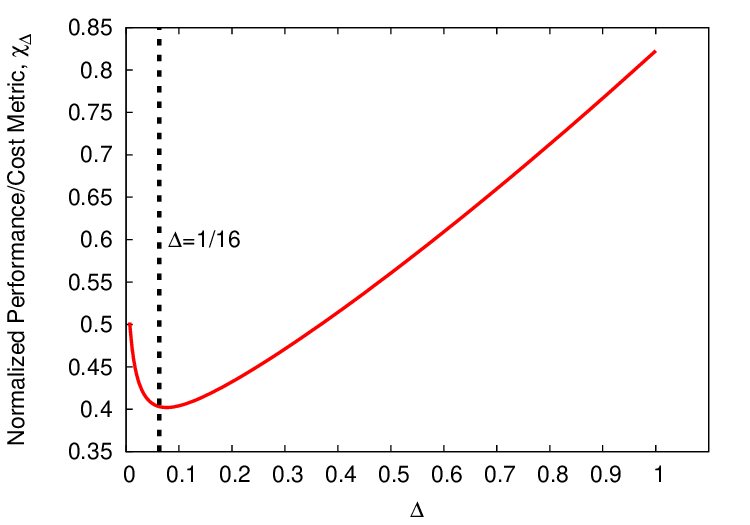}\tabularnewline
(\emph{b})\tabularnewline
\end{tabular}\end{center}

\begin{center}~\vfill\end{center}

\begin{center}\textbf{Fig. 11 - N. Anselmi} \textbf{\emph{et al.}}\textbf{,}
\textbf{\emph{{}``}}A Divide-and-Conquer Tiling Method ...''\end{center}

\newpage
\begin{center}~\vfill\end{center}

\begin{center}\begin{tabular}{c}
\includegraphics[%
  width=0.70\columnwidth]{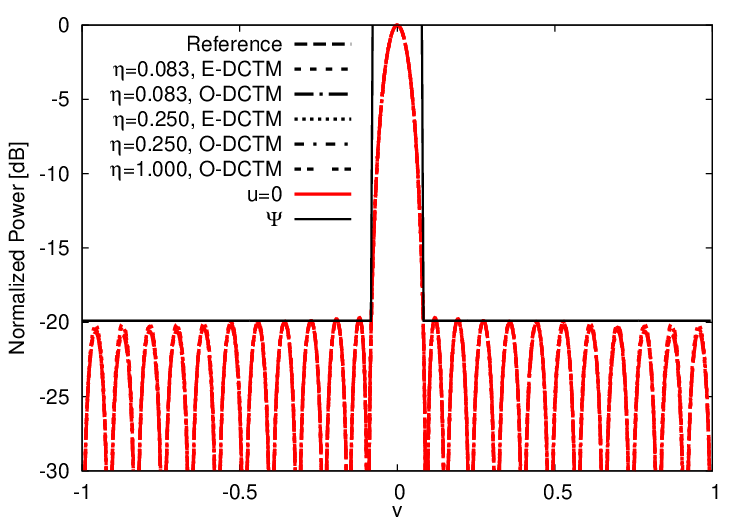}\tabularnewline
(\emph{a})\tabularnewline
\includegraphics[%
  width=0.70\columnwidth]{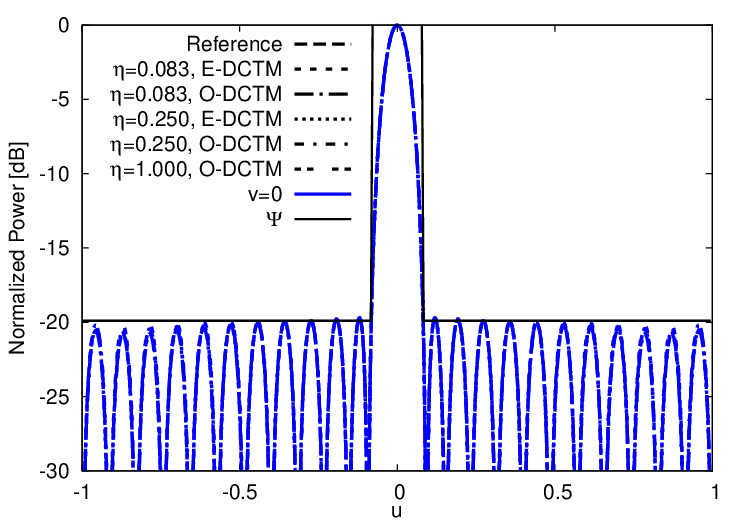}\tabularnewline
(\emph{b})\tabularnewline
\end{tabular}\end{center}

\begin{center}~\vfill\end{center}

\begin{center}\textbf{Fig. 12 - N. Anselmi} \textbf{\emph{et al.}}\textbf{,}
\textbf{\emph{{}``}}A Divide-and-Conquer Tiling Method ...''\end{center}

\newpage
\begin{center}~\vfill\end{center}

\begin{center}\begin{tabular}{ccc}
\includegraphics[%
  width=0.30\columnwidth]{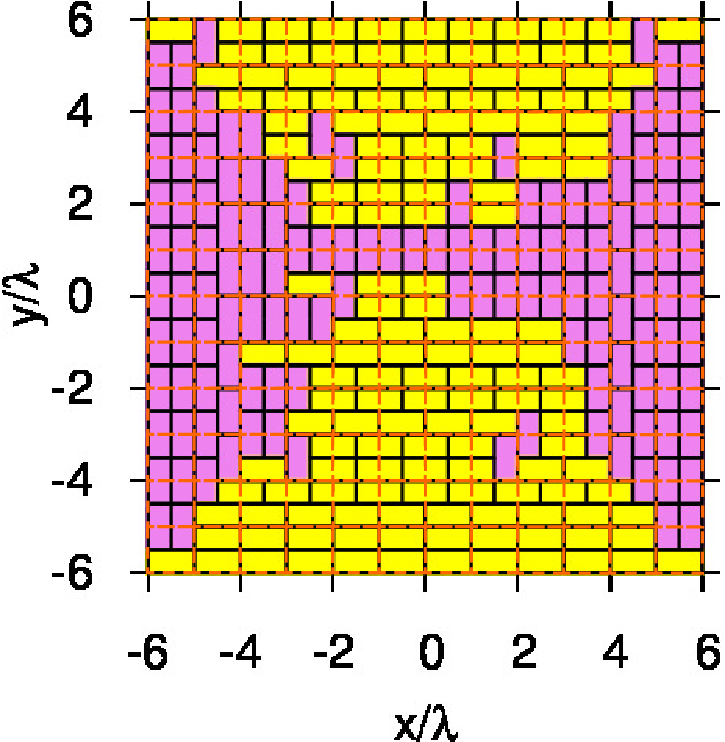}&
\includegraphics[%
  width=0.30\columnwidth]{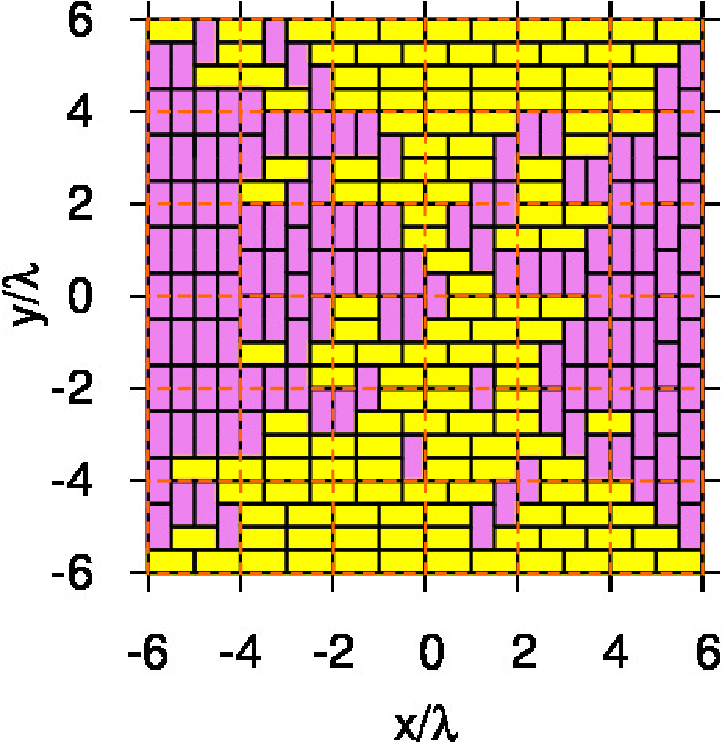}&
\includegraphics[%
  width=0.30\columnwidth]{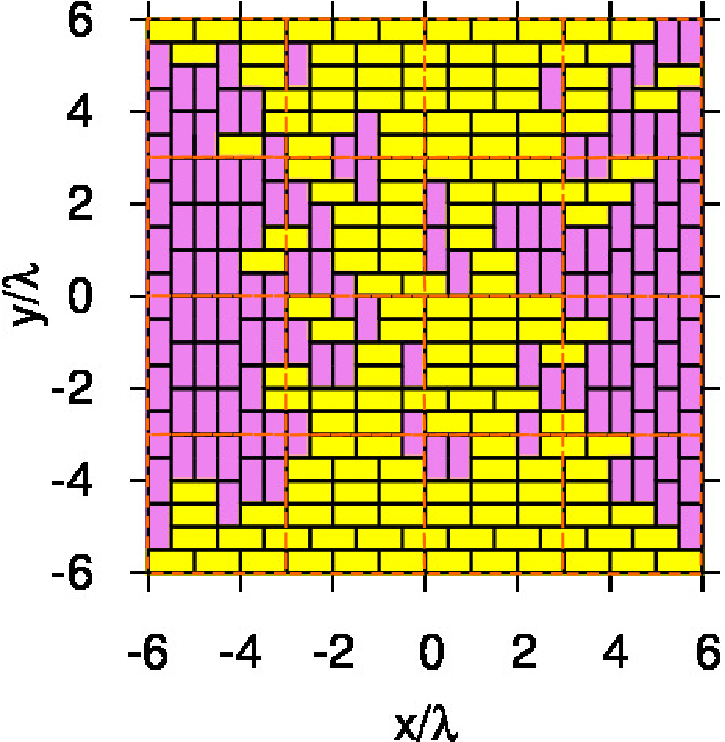}\tabularnewline
(\emph{a})&
(\emph{b})&
(\emph{c})\tabularnewline
&
&
\tabularnewline
\includegraphics[%
  width=0.30\columnwidth]{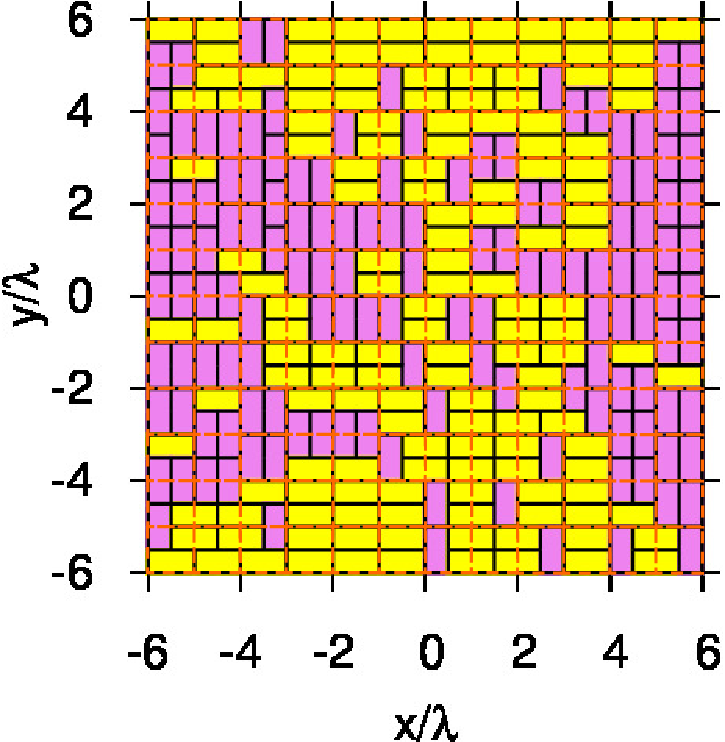}&
\includegraphics[%
  width=0.30\columnwidth]{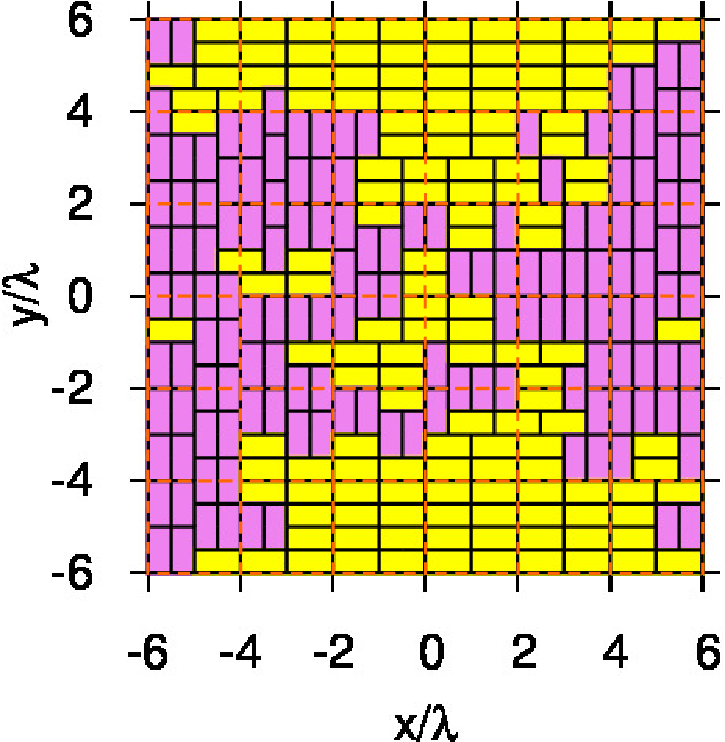}&
\includegraphics[%
  width=0.30\columnwidth]{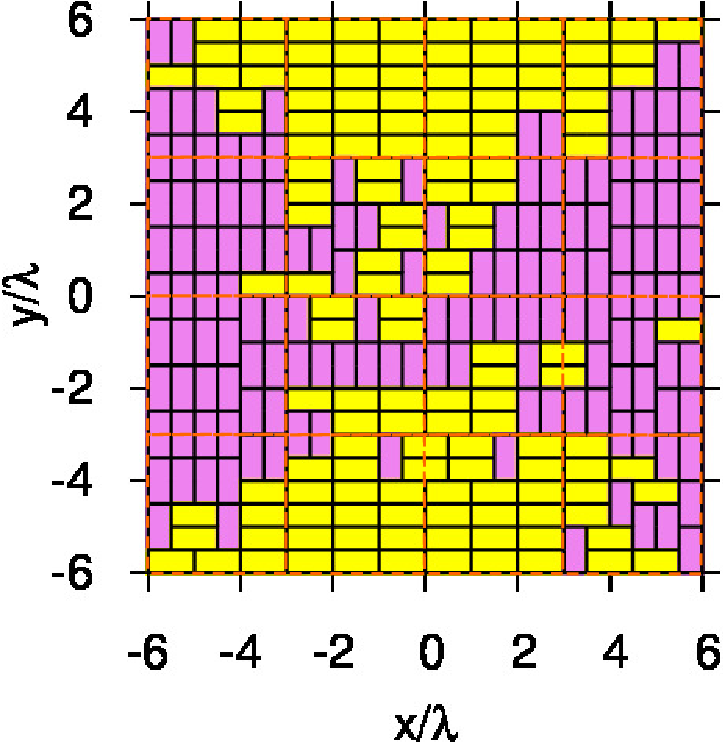}\tabularnewline
(\emph{d})&
(\emph{e})&
(\emph{f})\tabularnewline
\end{tabular}\end{center}

\begin{center}~\vfill\end{center}

\begin{center}\textbf{Fig. 13 - N. Anselmi} \textbf{\emph{et al.}}\textbf{,}
\textbf{\emph{{}``}}A Divide-and-Conquer Tiling Method ...''\end{center}

\newpage
\begin{center}~\vfill\end{center}

\begin{center}\begin{tabular}{c}
\includegraphics[%
  width=0.30\columnwidth]{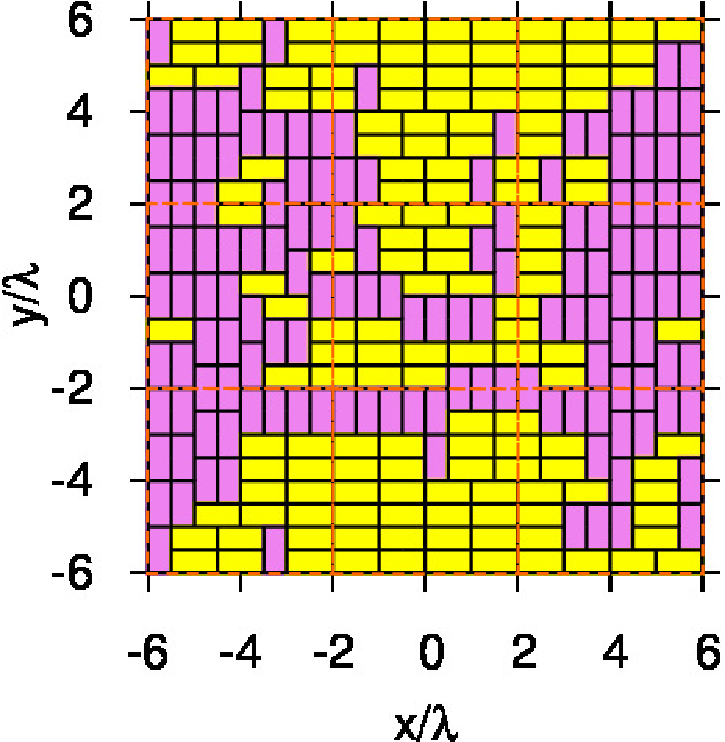}\tabularnewline
(\emph{a})\tabularnewline
\includegraphics[%
  width=0.30\columnwidth]{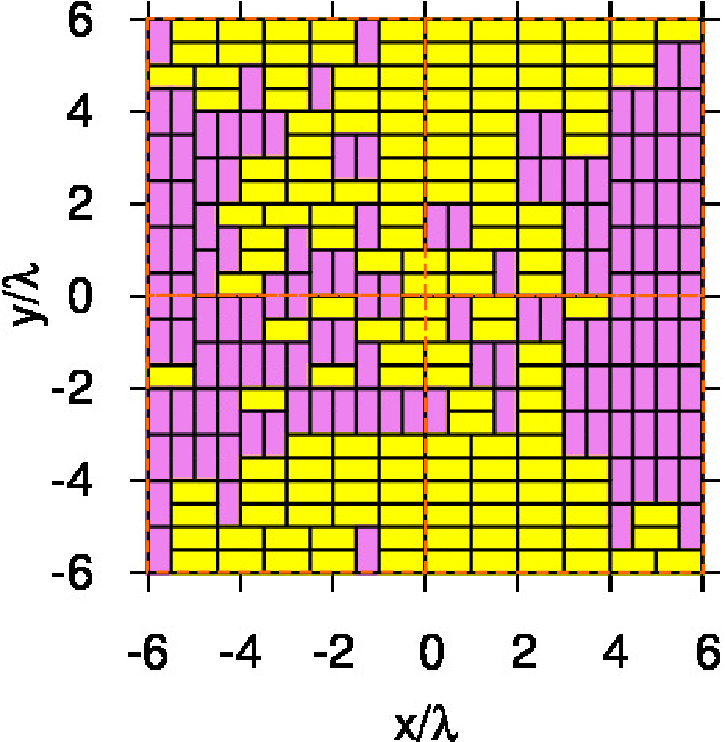}\tabularnewline
(\emph{b})\tabularnewline
\includegraphics[%
  width=0.30\columnwidth]{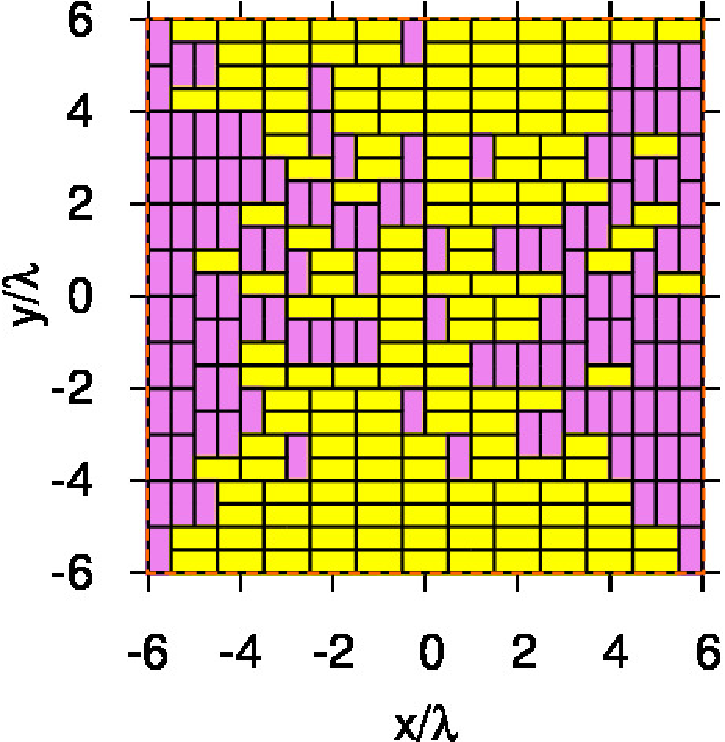}\tabularnewline
(\emph{c})\tabularnewline
\end{tabular}\end{center}

\begin{center}~\vfill\end{center}

\begin{center}\textbf{Fig. 14 - N. Anselmi} \textbf{\emph{et al.}}\textbf{,}
\textbf{\emph{{}``}}A Divide-and-Conquer Tiling Method ...''\end{center}

\newpage
\begin{center}~\vfill\end{center}

\begin{center}\begin{tabular}{c}
\includegraphics[%
  width=0.60\columnwidth]{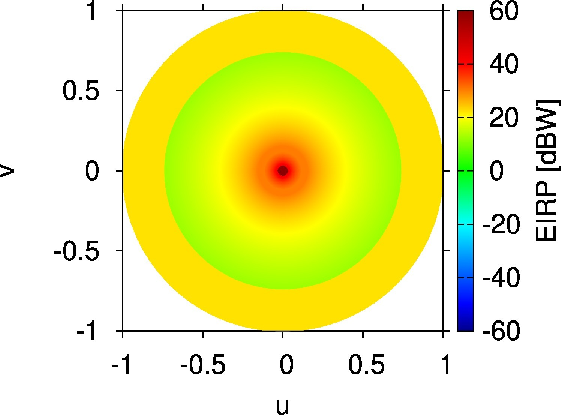}\tabularnewline
(\emph{a})\tabularnewline
\includegraphics[%
  width=0.60\columnwidth]{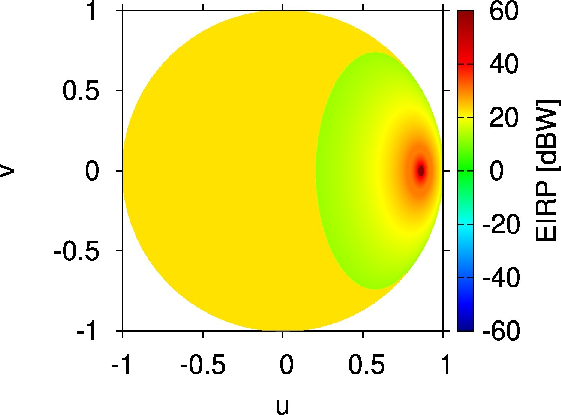}\tabularnewline
(\emph{b})\tabularnewline
\end{tabular}\end{center}

\begin{center}~\vfill\end{center}

\begin{center}\textbf{Fig. 15 - N. Anselmi} \textbf{\emph{et al.}}\textbf{,}
\textbf{\emph{{}``}}A Divide-and-Conquer Tiling Method ...''\end{center}

\newpage
\begin{center}~\vfill\end{center}

\begin{center}\begin{tabular}{cc}
\multicolumn{2}{c}{\includegraphics[%
  width=0.45\columnwidth]{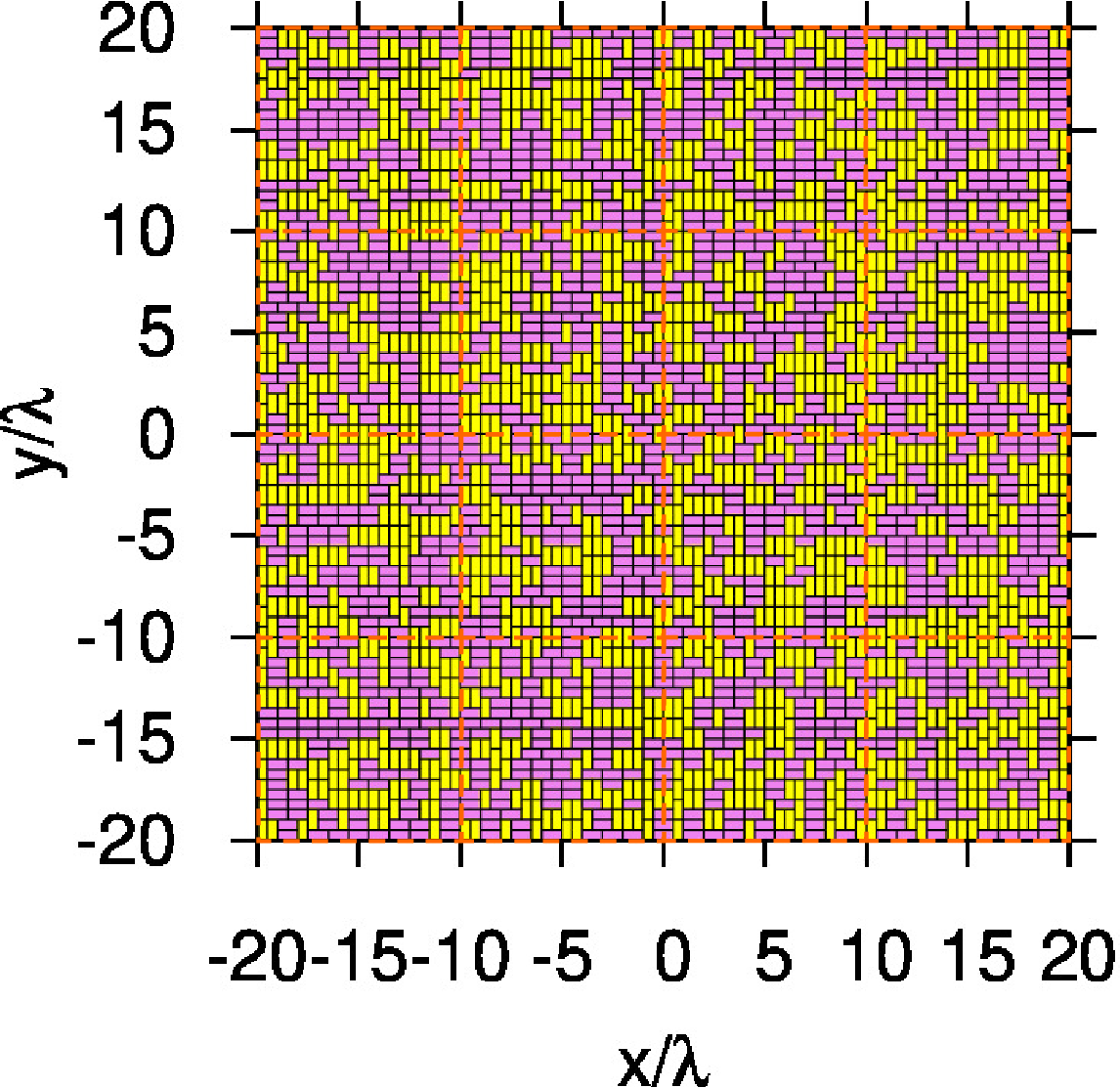}}\tabularnewline
\multicolumn{2}{c}{(\emph{a})}\tabularnewline
\includegraphics[%
  width=0.45\columnwidth]{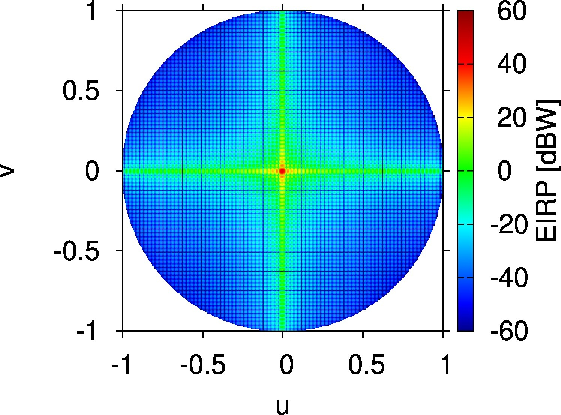}&
\includegraphics[%
  width=0.45\columnwidth]{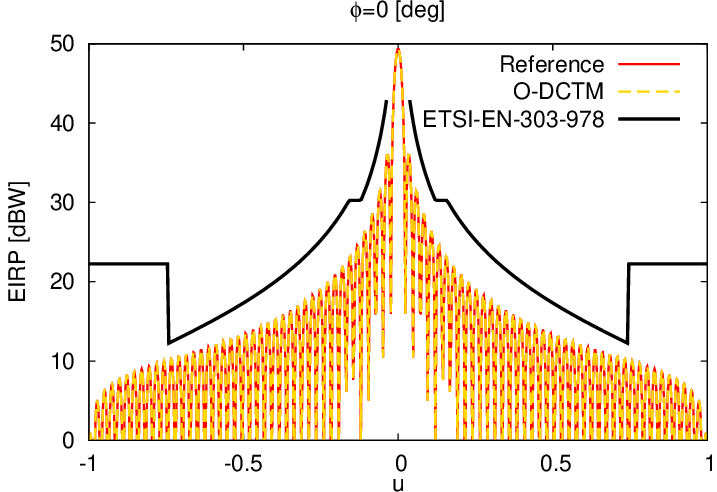}\tabularnewline
(\emph{b})&
(\emph{c})\tabularnewline
\end{tabular}\end{center}

\begin{center}~\vfill\end{center}

\begin{center}\textbf{Fig. 16 - N. Anselmi} \textbf{\emph{et al.}}\textbf{,}
\textbf{\emph{{}``}}A Divide-and-Conquer Tiling Method ...''\end{center}

\newpage
\begin{center}~\vfill\end{center}

\begin{center}\begin{tabular}{cc}
\includegraphics[%
  width=0.45\columnwidth]{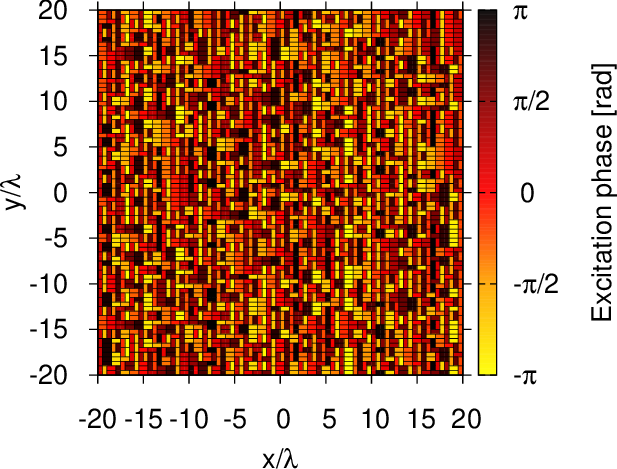}&
\includegraphics[%
  width=0.45\columnwidth]{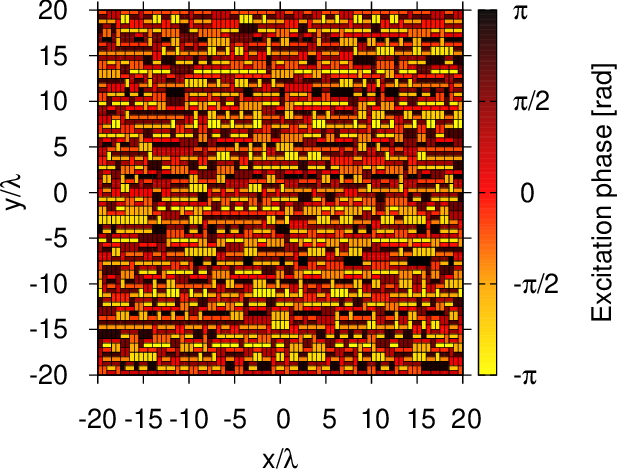}\tabularnewline
(\emph{a})&
(\emph{b})\tabularnewline
\includegraphics[%
  width=0.45\columnwidth]{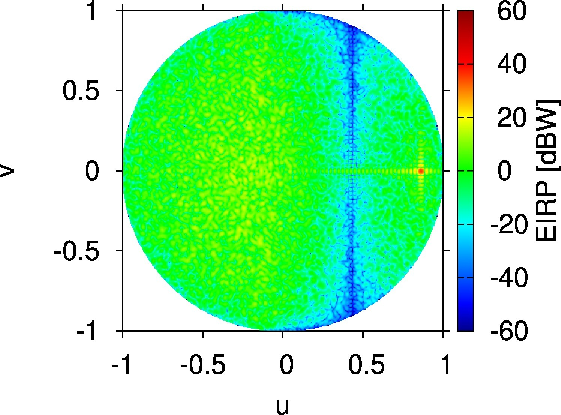}&
\includegraphics[%
  width=0.45\columnwidth]{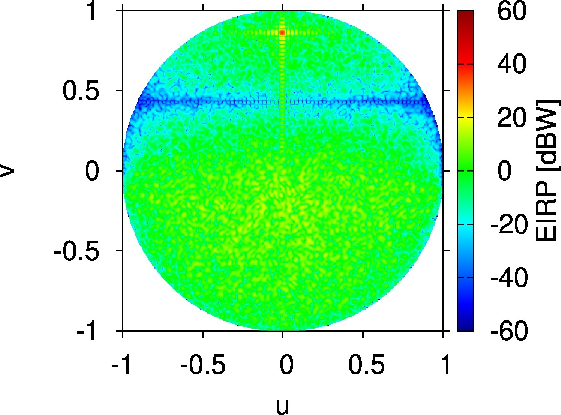}\tabularnewline
(\emph{c})&
(\emph{d})\tabularnewline
\includegraphics[%
  width=0.45\columnwidth]{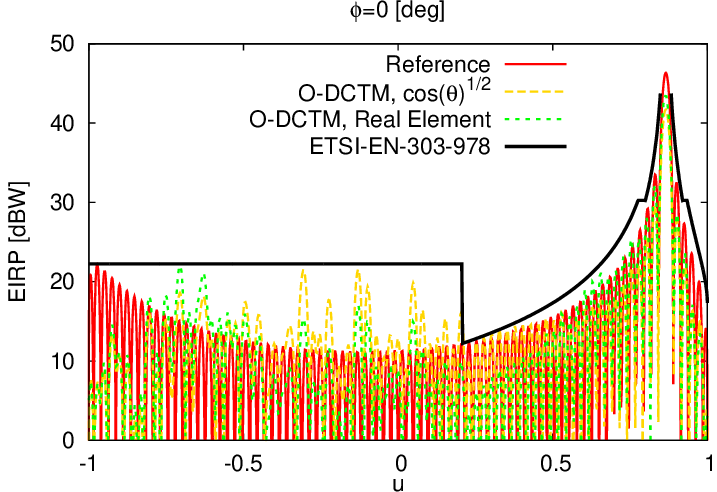}&
\includegraphics[%
  width=0.45\columnwidth]{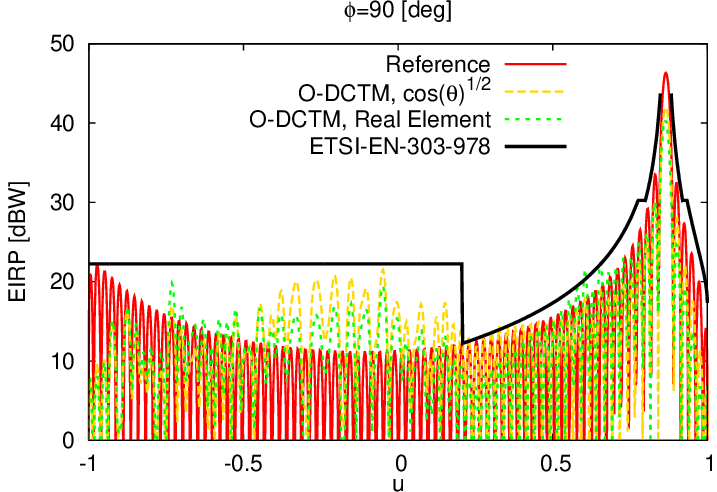}\tabularnewline
(\emph{e})&
(\emph{f})\tabularnewline
\end{tabular}\end{center}

\begin{center}~\vfill\end{center}

\begin{center}\textbf{Fig. 17 - N. Anselmi} \textbf{\emph{et al.}}\textbf{,}
\textbf{\emph{{}``}}A Divide-and-Conquer Tiling Method ...''\end{center}

\newpage
\begin{center}~\vfill\end{center}

\begin{center}\begin{tabular}{|c|c|c|c|c|c|}
\hline 
&
$SLL$ &
$D$ &
$HPBW_{az}$ &
$HPBW_{el}$ &
$\Phi$\tabularnewline
&
{[}dB{]}&
{[}dBi{]}&
{[}deg{]}&
{[}deg{]}&
\tabularnewline
\hline
\hline 
\emph{Reference}&
$-27.78$&
$21.60$&
$16.00$&
$16.00$&
$0.00$\tabularnewline
\hline 
\emph{ETM}&
$-21.24$&
$21.86$&
$15.44$&
$15.44$&
$1.14\times10^{-3}$\tabularnewline
\hline 
\emph{E-DCTM}&
$-20.73$&
$21.85$&
$15.53$&
$15.46$&
$1.17\times10^{-3}$\tabularnewline
\hline
\end{tabular}\end{center}

\begin{center}~\vfill\end{center}

\begin{center}\textbf{Tab. I - N. Anselmi} \textbf{\emph{et al.}}\textbf{,}
\textbf{\emph{{}``}}A Divide-and-Conquer Tiling Method ...''\end{center}

\newpage
\begin{center}~\vfill\end{center}

\begin{center}\begin{tabular}{|c|c|c|c|c|c|c|c|}
\hline 
&
$SLL$ &
$D$ &
$HPBW_{az}$ &
$HPBW_{el}$&
$\Phi$&
$T$&
\emph{$\tau$}\tabularnewline
&
{[}dB{]}&
{[}dBi{]}&
{[}deg{]}&
 {[}deg{]}&
&
&
{[}sec{]}\tabularnewline
\hline
\hline 
\emph{Reference}&
$-20.00$&
$28.46$&
$4.82$&
$9.13$&
$0.0$&
-&
-\tabularnewline
\hline
\emph{E-DCTM}&
$-19.10$&
$28.52$&
$4.82$&
$9.11$&
$7.77\times10^{-5}$&
$230$&
$74.6$\tabularnewline
\hline
{[}Anselmi 2017{]}&
$-19.32$&
$28.51$&
$4.82$&
$9.11$&
-&
$462\times10^{3}$&
$3.6\times10^{4}$\tabularnewline
\hline
{[}Yang 2021{]}&
$-19.32$&
$28.70$&
$4.82$&
$9.13$&
-&
-&
$108.8$\tabularnewline
\hline
\end{tabular}\end{center}

\begin{center}~\vfill\end{center}

\begin{center}\textbf{Tab. II - N. Anselmi} \textbf{\emph{et al.}}\textbf{,}
\textbf{\emph{{}``}}A Divide-and-Conquer Tiling Method ...''\end{center}

\newpage
\begin{center}~\vfill\end{center}

\begin{center}\begin{tabular}{|c|c|c|c|c|c|c|c|}
\hline 
$\widehat{M}\times\widehat{N}$&
$\eta$&
$SLL$ &
$D$&
$HPBW_{az}$ &
$HPBW_{el}$ &
$\Phi$&
$T$\tabularnewline
&
&
{[}dB{]}&
{[}dBi{]}&
{[}deg{]}&
{[}deg{]}&
{[}$\times10^{-6}]$&
\tabularnewline
\hline
\hline 
\multicolumn{8}{|c|}{\emph{Reference}}\tabularnewline
\hline
\hline 
-&
-&
$-20.00$&
$31.59$&
$4.43$&
$4.43$&
$0.00$&
-\tabularnewline
\hline
\hline 
\multicolumn{8}{|c|}{\emph{E-DCTM}}\tabularnewline
\hline
\hline 
$2\times2$&
$\frac{1}{12}$&
$-19.71$&
$31.63$&
$4.43$&
$4.43$&
$4.55$&
$448$\tabularnewline
\hline 
$3\times3$&
$\frac{1}{8}$&
$-19.73$&
$31.63$&
$4.43$&
$4.43$&
$4.74$&
$768$\tabularnewline
\hline 
$4\times4$&
$\frac{1}{6}$&
$-19.73$&
$31.63$&
$4.43$&
$4.43$&
$4.91$&
$3672$\tabularnewline
\hline 
$6\times6$&
$\frac{1}{4}$&
$-19.71$&
$31.63$&
$4.43$&
$4.43$&
$4.90$&
$802115$\tabularnewline
\hline
\hline 
\multicolumn{8}{|c|}{\emph{O-DCTM}}\tabularnewline
\hline
\hline 
$2\times2$&
$\frac{1}{12}$&
$-19.69$&
$31.66$&
$4.43$&
$4.43$&
$7.34$&
$432$\tabularnewline
\hline 
$3\times3$&
$\frac{1}{8}$&
$-19.69$&
$31.64$&
$4.43$&
$4.43$&
$6.28$&
$509$\tabularnewline
\hline 
$4\times4$&
$\frac{1}{6}$&
$-19.69$&
$31.66$&
$4.43$&
$4.43$&
$5.79$&
$1307$\tabularnewline
\hline 
$6\times6$&
$\frac{1}{4}$&
$-19.72$&
$31.64$&
$4.43$&
$4.43$&
$5.09$&
$2858$\tabularnewline
\hline 
$8\times8$&
$\frac{1}{3}$&
$-19.69$&
$31.65$&
$4.43$&
$4.43$&
$5.76$&
$5020$\tabularnewline
\hline 
$12\times12$&
$\frac{1}{2}$&
$-19.72$&
$31.64$&
$4.43$&
$4.43$&
$4.97$&
$11889$\tabularnewline
\hline 
$24\times24$&
$1$&
$-19.72$&
$31.64$&
$4.43$&
$4.43$&
$4.74$&
$99161$\tabularnewline
\hline
\end{tabular}\end{center}

\begin{center}~\vfill\end{center}

\begin{center}\textbf{Tab. III - N. Anselmi} \textbf{\emph{et al.}}\textbf{,}
\textbf{\emph{{}``}}A Divide-and-Conquer Tiling Method ...''\end{center}

\newpage
\begin{center}~\vfill\end{center}

\begin{center}\begin{tabular}{|c|c|c|c|c|c|c|}
\hline 
$\left(\theta_{0},\phi_{0}\right)$ &
$SLL$ &
$D$ &
$EIRP$&
$HPBW_{az}$ &
$HPBW_{el}$&
$\Phi$\tabularnewline
{[}deg{]}&
{[}dB{]}&
{[}dBi{]}&
{[}dBW{]}&
{[}deg{]}&
 {[}deg{]}&
\tabularnewline
\hline
\hline 
\multicolumn{7}{|c|}{\emph{Reference}}\tabularnewline
\hline
\hline 
$\left(0,0\right)$&
$-13.30$&
$43.37$&
$49.39$&
$1.22$&
$1.22$&
$0.00$\tabularnewline
\hline 
$\left(60,0\right)$&
$-13.24$&
$40.32$&
$46.34$&
$2.45$&
$1.22$&
$0.00$\tabularnewline
\hline 
$\left(60,90\right)$&
$-13.24$&
$40.32$&
$46.34$&
$1.22$&
$2.45$&
$0.00$\tabularnewline
\hline
\hline 
\multicolumn{7}{|c|}{\emph{O-DCTM}}\tabularnewline
\hline
\hline 
$\left(0,0\right)$&
$-13.30$&
$43.37$&
$49.39$&
$1.22$&
$1.22$&
$0.00$\tabularnewline
\hline 
$\left(60,0\right)$&
$-12.52$&
$35.69$&
$41.71$&
$2.52$&
$1.27$&
$7.5\times10^{-12}$\tabularnewline
\hline
$\left(60,90\right)$&
$-12.59$&
$35.95$&
$41.97$&
$1.28$&
$2.53$&
$7.7\times10^{-12}$\tabularnewline
\hline
\end{tabular}\end{center}

\begin{center}~\vfill\end{center}

\begin{center}\textbf{Tab. IV - N. Anselmi} \textbf{\emph{et al.}}\textbf{,}
\textbf{\emph{{}``}}A Divide-and-Conquer Tiling Method ...''\end{center}
\end{document}